\documentclass[draft]{agujournal}

\draftfalse

\usepackage{amsmath}  

\journalname{JGR-Planets}

\newcommand{\mitbf}[1]{
  \hbox{\mathversion{bold}$#1$}}

\begin{document}

%
%


\title{The Cassini State of the Moon's inner core}

%
%




 \authors{
Christopher Stys \affil{1} and Mathieu Dumberry \affil{1}}
\affiliation{1}{Department of Physics, University of Alberta, Edmonton, Alberta, Canada.}






\correspondingauthor{Mathieu Dumberry}{dumberry@ualberta.ca}




\begin{keypoints}
\item Three Cassini states associated with the inner core of the Moon are possible and depend on the free inner core nutation (FICN) frequency.
\item Because the FICN frequency is close to the Lunar precession frequency, resonant amplification can lead to a large inner core tilt angle.
\item The inner core can be misaligned from the mantle by as much as 33 degrees towards the orbit normal, or 17 degrees away from it.
\end{keypoints}

%
%




\begin{abstract}
We present a model of the precession dynamics of the Moon that comprises a fluid outer core and a solid inner core.  We show that three Cassini states associated with the inner core exist.  The tilt angle of the inner core in each of these states is determined by the ratio between the free inner core nutation frequency ($\omega_{ficn}$) and the precession frequency $\Omega_p = 2\pi/18.6$ yr $^{-1}$.   All three Cassini states are possible if $|\omega_{ficn}| > 2\pi/16.4$ yr $^{-1}$, but only one is possible otherwise.  Assuming that the lowest energy state is favoured, this transition marks a discontinuity in the tilt  angle of the inner core, transiting from $-33^\circ$ to $17^\circ$ as measured with respect to the mantle figure axis, where negative angles indicate a tilt towards the orbit normal. Possible Lunar interior density structures cover a range of $\omega_{ficn}$, from approximately half to twice as large as $\Omega_p$, so the precise tilt angle of the inner core remains unknown, though it is likely large because $\Omega_p$ is within the resonant band of $\omega_{ficn}$.   Adopting one specific density model, we suggest an inner core tilt of approximately $-17^\circ$.  Viscoelastic deformations within the inner core and melt and growth at the surface of a tilted inner core, both neglected in our model, should reduce this amplitude.  If the inner core is larger than approximately 200 km, it may contribute by as much as a few thousandths of a degree on the observed mantle precession angle of $1.543^\circ$.  
\end{abstract}

\section{Introduction}
\label{sec:intro}

\begin{figure}
\begin{center}
\includegraphics[width=10cm]{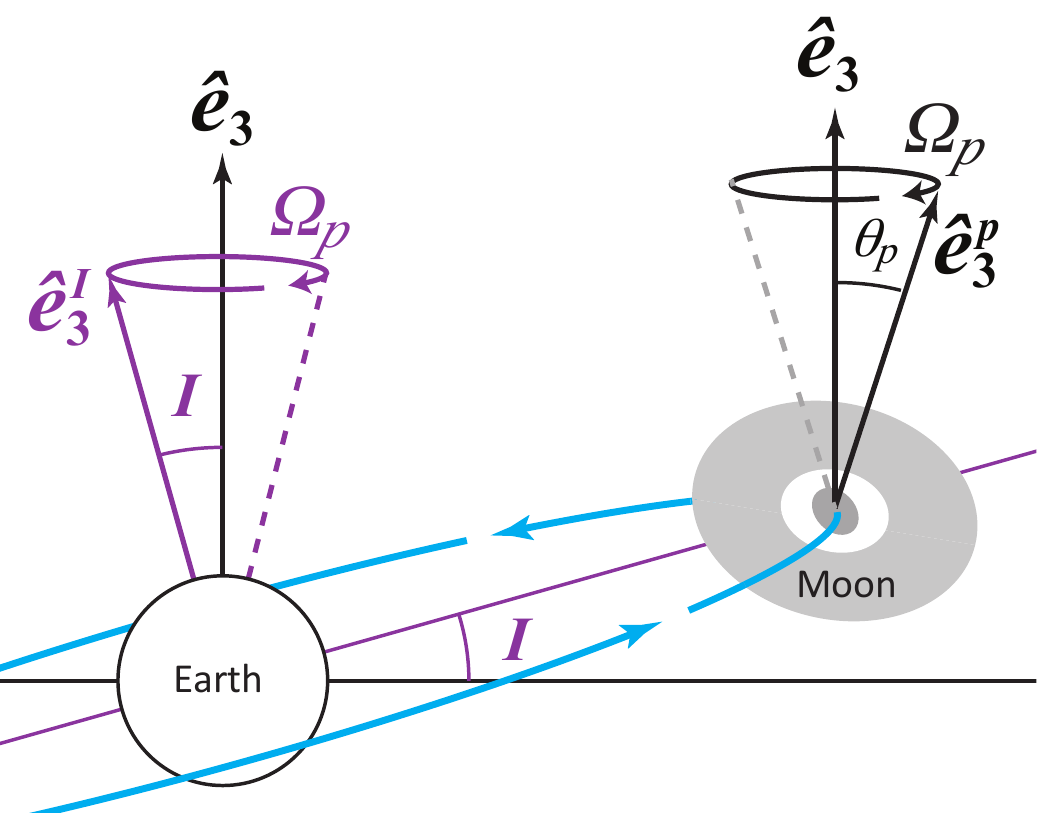}
\caption{\label{fig:orb} The Earth-Moon orbital dynamics. The plane of the Moon's orbit around the Earth (light blue) is inclined by an angle $I=5.145^\circ$ with respect to the ecliptic normal (pointing in direction $\mitbf{\hat{e}_3}$).  The orientation of the orbital plane (normal vector pointing in direction $\mitbf{\hat{e}^I_3}$) precesses in a retrograde direction at a frequency of $\Omega_p=2\pi/18.6$ yr$^{-1}$ about $\mitbf{\hat{e}_3}$.  The symmetry axis of the Moon's mantle (pointing in direction $\mitbf{\hat{e}^p_3}$) is inclined by $\theta_p=1.543^\circ$ with respect to the ecliptic normal, in the same plane as $I$ but in the opposite direction, and is also precessing at frequency $\Omega_p$. }
\end{center}
\end{figure}

The Moon's orbital position and orientation in space have been extensively studied by Lunar Laser Ranging (LLR) \cite[][]{dickey94}.  The Moon is in a tidally locked 1:1 spin-orbit resonance around the Earth, with its orbital period of 27.322 days equal to its rotation period.  The Lunar orbital plane is inclined by $I = 5.145^{\circ}$ with respect to the ecliptic plane and the spin-symmetry axis of the mantle is inclined by an angle $\theta_p=1.543^{\circ}$, in the opposite direction of $I$ (Fig.~\ref{fig:orb}).  The spin-symmetry axis and the normal to the orbital plane remain co-planar with, and are precessing about, the normal to the ecliptic in a retrograde direction with a frequency of $\Omega_p = 2 \pi /18.6$ yr $^{-1}$.  The latter configuration describes a Cassini state \cite[][]{colombo66,peale69}.   Observations suggest that the Moon is slightly offset from an exact Cassini state by a small angle of 0.26 arcsec, indicating the presence of a dissipation mechanism \cite[][]{yoder81,williams01}. 

LLR observations only reveal the orientation of the solid outer region of the Moon made up of its mantle and crust.  Several lines of evidence suggest that the Moon has a small fluid metallic core \cite[][]{wieczorek06},  with perhaps a solid inner core at its centre \citep{weber11}.  As shown in \cite{dumberry16} (hereinafter referred to as DW16), both the spin axis of the fluid core and the spin-symmetry axis of the solid inner core are expected to undergo a retrograde 18.6 yr precession and be part of the Cassini state of the Moon.  Whether they remain aligned with the mantle depends on the frequency of two free precession modes, the free core nutation (FCN) and free inner core nutation (FICN).  The FCN describes the free precession of the spin axis of the fluid core when it is displaced from an alignment with the mantle figure axis.  Similarly,  the FICN describes the free precession of the spin-symmetry axis of the inner core when it is misaligned from the mantle.  For the Moon, both of these free modes are retrograde.  If the FCN frequency is much faster than the forcing precession frequency $\Omega_p$, the spin vector of the fluid core should remain closely aligned with the mantle spin-symmetry axis.  Likewise, if the frequency of the FICN mode is much faster than $\Omega_p$, the inner core should be gravitationally locked to the mantle and nearly aligned with it. These free modes have not been observed directly so are not known, but their frequencies depend on the interior density structure.   

The period of the FCN, as seen in the inertial frame, is estimated to be longer than 150 yrs \cite[e.g.][]{gusev08,williams14b}, implying that the fluid core is not efficiently entrained by the 18.6 yr mantle precession. The rotation vector of the fluid core should therefore remain in close alignment with the normal to the ecliptic \cite[e.g.][]{poincare10,goldreich67,meyer11}.  For models of the Moon's interior compatible with geodetic and seismic constraints, the period of the FICN mode (in the inertial frame) is estimated to be in the range of a few years to a few decades (DW16).  The 18.6 yr forcing period is therefore within the resonance band of the FICN.   As a consequence, the precession angle of the inner core is very sensitive to the chosen density and elliptical structure of the Moon and may be largely misaligned from the mantle.

A model to compute the Cassini state of a Moon that comprises a fluid core and solid inner core was presented in DW16.  The model was developed under the assumption of small angles of precession.  Although this is suitable for the present-day mantle tilt of $1.543^{\circ}$, it is not valid in the past when the Moon was closer to Earth and the mantle tilt might have been as high as $49^{\circ}$ \cite[][]{ward75}. Furthermore, because of the proximity of the forcing period to being in resonance with the FICN, the inner core could have a relatively large angle of misalignment with the mantle.  

The objective of the present study is to develop a model of the internal Cassini state of the Moon that is more general than the one developed in DW16, one which remains valid for large angles of misalignments. Besides the general aim of furthering our understanding of the precession dynamics of planetary bodies, the motivation of our study is also to lay the foundation for two planned applications.  

First, whether the Moon has a solid inner core remains uncertain.  Thermal evolution models suggest that a solid inner core should have crystallized at its centre \citep{zhang13,laneuville14,scheinberb15}.  A solid inner core with a radius of $240 \pm 10$ km has been inferred based on seismic data \citep{weber11} but this interpretation is not unique \cite[e.g.][]{garcia11}.  If an inner core is present, it may be possible to detect it through its precession dynamics. As seen in the mantle frame, a misaligned inner core is precessing with a period of 1 Lunar day, causing a periodic variation in the degree 2 order 1 coefficients of gravity \citep{williams07}, which may be detectable \cite[e.g.][]{zuber13}.  The amplitude of this signal depends on the size of the inner core but also on its angle of precession. Thus, it is important to compute this angle correctly.

Second, as mentioned above, in the past, when the Moon was closer to Earth, the mantle tilt angle was much larger than the current $1.543^{\circ}$.  The misalignment between the spin axis of the fluid core and the mantle, and thus the differential velocity at the core-mantle boundary, was possibly sufficiently large to have generated a dynamo \cite[][]{williams01,dwyer11}.  Furthermore, the differential velocity at the inner core boundary may have also been higher in the past, also perhaps large enough to generate dynamo action.   To assess these possibilities, it is necessary to construct a model of the precession dynamics of the Moon that remains valid for large angles.

\section{Theory} 
 
The rotational model of the Moon that we develop below is based on the model presented in DW16,  which is itself an adaptation of a model developed to study Earth's nutations.  The original nutation model is presented in detail in \cite{mathews91a}.  

The procedure that we follow is, first, to define a reference interior model of the Moon (section 2.1).  This interior model is constructed under the assumption that no external torque acts on the Moon.  We then place this reference model in orbit about Earth, subject to its gravitational field, and consider how the alignment of the symmetry axes and rotation vectors of each region is altered in the Cassini state.  To do so, we must properly define each of these vectors in the reference frame attached to the rotating mantle, the frame in which the nutation model of \cite{mathews91a} is developed.  This is done in section 2.2.  The rotational model is then developed in section 2.3.

\subsection{The interior density model of the Moon}

We assume a simple model of the Moon of mass $M$ with an external radius $R$, a solid inner core of radius $r_s$, a fluid outer core of radius $r_f$, a crust of thickness $h_c$, and a mantle with an outer radius of $r_m=R-h_c$. The densities of the solid inner core ($\rho_s$), fluid core ($\rho_f$), mantle ($\rho_m$) and crust ($\rho_c$) are assumed uniform.  Adopting uniform density layers amounts to neglecting compressibility effects from increasing pressure with depth.  Given the small pressures in the Moon's interior (less than about 5 GPa), this is a good first order description.  

The precession model that we develop below involves the principal moments of inertia of each region.  The latter are related to the spherical harmonic degree two coefficients of the gravity field of the Moon.  For convenience, we assume a reference model in which the principal moments of inertia of each region are aligned.  Although in reality this is unlikely to be the case because the surface topography of degree two is not aligned with the degree two gravity field \cite[e.g.][]{araki09,smith10}, this assumption greatly simplifies our reference model.  Since we assume uniform density layers, all contributions to the non-spherical gravity field (i.e. all mass anomalies) are caused by topography at region boundaries.  The principal moments of inertia of each region are then connected to the degree two topography at region boundaries, more specifically to the polar and equatorial flattening.  We define the polar flattening as the difference between the equatorial and polar radius, divided by the mean spherical radius.  Likewise, we define the equatorial flattening as the difference between the maximum and minimum equatorial radius, divided by the mean spherical radius. We denote the polar flattening at the inner core boundary (ICB), core-mantle boundary (CMB), crust-mantle boundary and surface by $\epsilon_s$, $\epsilon_f$, $\epsilon_m$, and $\epsilon_r$, respectively.  The difference between the equatorial and polar radius at each of these interfaces is then $r_s \epsilon_s$, $r_f \epsilon_f$, $r_m \epsilon_m$, and $R \epsilon_r$, respectively.  The equatorial flattening at the same boundaries are denoted by $\xi_s$, $\xi_f$, $\xi_m$ and $\xi_r$, respectively.  The difference between the maximum and minimum equatorial radius at each of these interfaces is then $r_s \xi_s$, $r_f \xi_f$, $r_m \xi_m$, and $R \xi_r$, respectively.

The polar and equatorial flattenings of each region are connected to the principal moments of inertia of the whole Moon ($C>B>A$), fluid core ($C_f>B_f>A_f$) and solid inner core ($C_s>B_s>A_s$).  In particular, they are connected to the degree two coefficients of the gravity potential $J_2$ and $C_{22}$ by  

\begin{subequations}
\begin{align}
J_2 & = \frac{C-\bar{A}}{M R^2} = \frac{8\pi}{15} \frac{1}{M R^2} \left[ (\rho_s-\rho_f) r_s^5 \epsilon_{s}  + (\rho_f-\rho_m) r_f^5 \epsilon_f + (\rho_m-\rho_c) r_m^5 \epsilon_m + \rho_c R^5 \epsilon_r  \right] \, , \label{eq:j2} \\
C_{22} & = \frac{B-A}{4 M R^2} =\frac{8\pi}{15} \frac{1}{4 M R^2} \left[ (\rho_s-\rho_f) r_s^5 \xi_s + (\rho_f-\rho_m) r_f^5 \xi_f + (\rho_m-\rho_c) r_m^5 \xi_m + \rho_c R^5 \xi_r  \right] \, . \label{eq:c22} 
\end{align}
\label{eq:j2c22}
\end{subequations} 
where $\bar{A}$ is the mean equatorial moment of inertia of the whole Moon.  The latter, and the mean equatorial moments of the fluid core ($\bar{A}_f$) and inner core ($\bar{A}_s$) are defined as

\begin{equation}
\bar{A} = \frac{1}{2} (A+B)  \, , \hspace*{0.5cm} \bar{A}_f = \frac{1}{2} (A_f+B_f)  \,,  \hspace*{0.5cm} \bar{A}_s = \frac{1}{2} (A_s+B_s)   \, . \label{ref:meanH}
\end{equation}
From these, we define the dynamical ellipticities of the whole Moon ($e$), fluid core ($e_f$) and solid inner core ($e_s$), 

\begin{equation}
e = \frac{C-\bar{A}}{\bar{A}} \, \hspace*{0.5cm} e_f = \frac{C_f-\bar{A}_f}{\bar{A}_f} \, \hspace*{0.5cm} e_s = \frac{C_s-\bar{A}_s}{\bar{A}_s} \, .
\label{eq:e}
\end{equation}
These dynamical ellipticities are important parameters of our model and the way in which they are calculated is explained in more details in section 3.1.

Although the density discontinuity at the crust-mantle boundary is taken into account in the interior mass distribution, the solid outer shell region that comprises both the crust and mantle constitute a single body in terms of the rotational dynamics.  For short, in the development of the rotational model below, we will refer to this outer region as the mantle. If we define the direction of the figure axis of this ``mantle'' by $\mitbf{\hat{e}^p_3}$, then the ellipsoidal figures of each region of our reference model are aligned, and are in uniform rotation at the sidereal frequency $2\pi/27.322$ day$^{-1}$ about $\mitbf{\hat{e}^p_3}$.

\subsection{Definition of the reference frames, symmetry axes and rotation vectors}

The focus of our study is to describe the equilibrium Cassini state of the Moon.  As such, we focus on the long timescale dynamics, and only consider the response of the Moon to the gravitational torque by Earth averaged over one orbit.  In other words, we neglect the modulation of the torque over one orbit and the small latitudinal and longitudinal librations of the Moon that result from it. This assumption is implicit in the presentation of our model and in the discussion of all our results.

To describe the Cassini state of the Moon, we must first define the possible reference frames in which to view the orbital and rotational dynamics.  We use three different reference frames in our study.  The first is the inertial reference frame, defined by unit vectors ($\mitbf{\hat{e}_1}$, $\mitbf{\hat{e}_2}$, $\mitbf{\hat{e}_3}$), with $\mitbf{\hat{e}_3}$ aligned with the ecliptic normal.  The second is a reference frame attached to the rotating mantle, defined by unit vectors $(\mitbf{\hat{e}^p_1}, \mitbf{\hat{e}^p_2}, \mitbf{\hat{e}^p_3})$.  We have already defined $\mitbf{\hat{e}^p_3}$ to be aligned with the maximum (polar) moment of inertia of the mantle.  $\mitbf{\hat{e}^p_1}$ and $\mitbf{\hat{e}^p_2}$ are aligned, respectively, with the minimum and intermediate moments of inertia (both in equatorial directions).  This is the frame in which we develop our dynamical model.  As mentioned in the introduction, the Cassini state is characterized by a tilt of $\mitbf{\hat{e}^p_3}$ from $\mitbf{\hat{e}_3}$, though both remain co-planar with the orbit normal ($\mitbf{\hat{e}^I_3}$). It is convenient to refer to the plane which contains all three as the ``Cassini plane''.   Viewed in the inertial frame, the Cassini plane is rotating in the retrograde direction at frequency $\Omega_p$ about an axis aligned with the ecliptic normal ($\mitbf{\hat{e}_3}$).  A third reference frame in which to view the rotational dynamics is then one attached to this Cassini plane.  We refer to this reference frame as the Cassini frame, defined by unit vectors ($\mitbf{\hat{e}^c_1}$, $\mitbf{\hat{e}^c_2}$, $\mitbf{\hat{e}^c_3}$).  Direction $\mitbf{\hat{e}^c_3}$ is aligned with the ecliptic normal, and the Cassini plane coincides with the surface defined by $\mitbf{\hat{e}^c_1}$ and $\mitbf{\hat{e}^c_3}$.  Direction $\mitbf{\hat{e}^c_2}$ is perpendicular to the Cassini plane and is aligned with the line of the descending node of the lunar orbit on the ecliptic plane.  More information about these three reference frames is given in Appendix A, where we also present the mathematical relationships that unite them.

\begin{figure}
\begin{center}
\includegraphics[width=13cm]{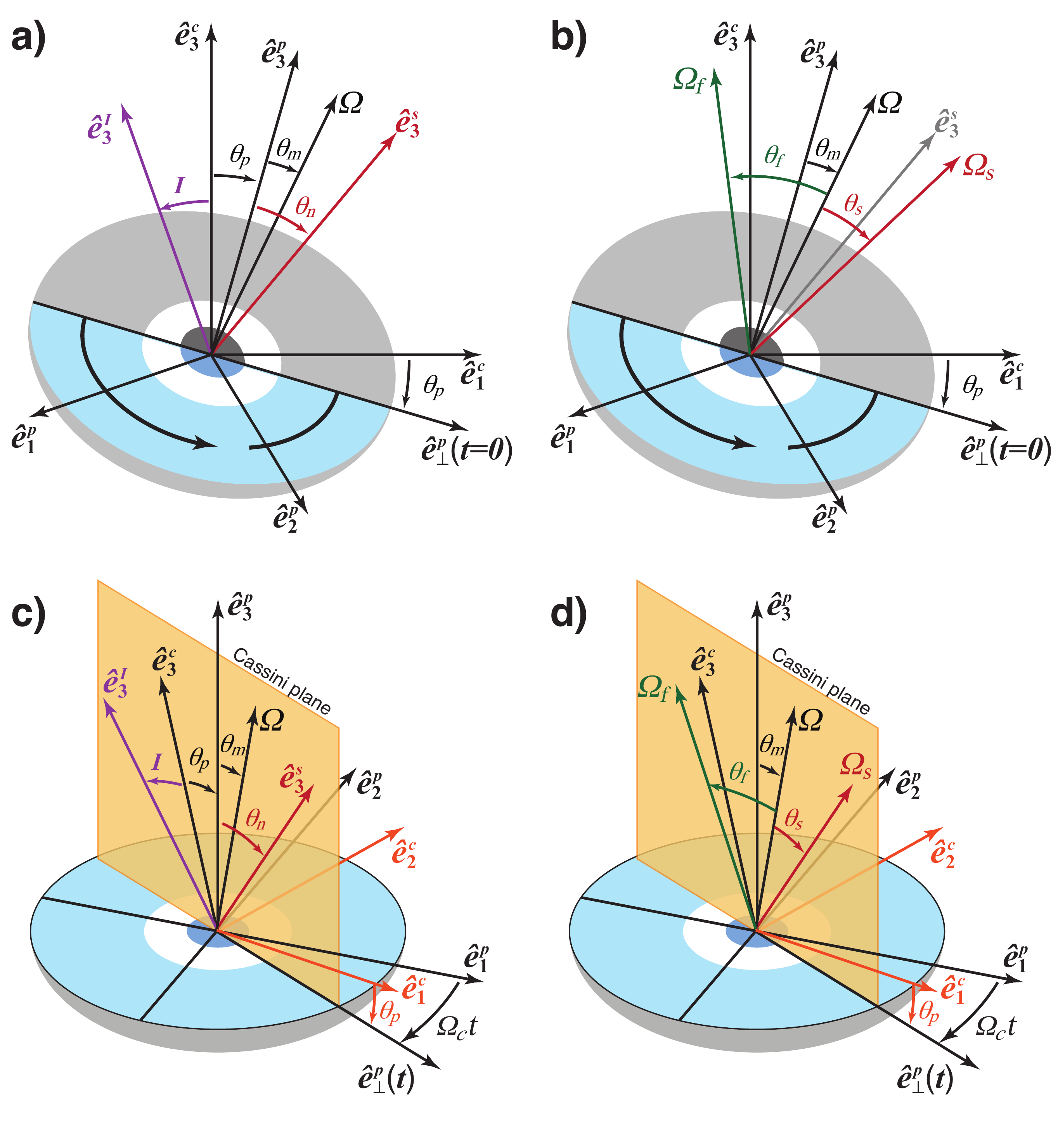}  
\caption{\label{fig:rotframes} 
The Cassini state of the Moon viewed (a, b) in the Cassini frame and (c, d) in a frame attached to the rotating mantle.  The Cassini frame is defined by unit vectors ($\mitbf{\hat{e}^c_1}$, $\mitbf{\hat{e}^c_2}$, $\mitbf{\hat{e}^c_3}$), the mantle frame by unit vectors ($\mitbf{\hat{e}^p_1}$, $\mitbf{\hat{e}^p_2}$, $\mitbf{\hat{e}^p_3}$).  Viewed in the Cassini frame (a, b), the orbit normal ($\mitbf{\hat{e}_3^I}$), the symmetry axes of the mantle ($\mitbf{\hat{e}^p_3}$) and inner core ($\mitbf{\hat{e}^s_3}$), and the rotation vectors of the mantle ($\mitbf{\Omega}$), fluid core ($\mitbf{\Omega_f}$) and inner core ($\mitbf{\Omega_f}$) remain at fixed orientations.   The light grey, white, and dark grey ellipsoid in panels (a) and (b) represent a polar cross-section of the mantle, fluid core and inner core, respectively.  Blue shaded parts show the equatorial cross section.  The black curved arrow in the equatorial plane of panels (a) and (b) indicates the direction of rotation, at frequency $\Omega_c$, of the mantle frame axes $\mitbf{\hat{e}^p_1}$ and $\mitbf{\hat{e}^p_2}$ about $\mitbf{\hat{e}^p_3}$.  Viewed in the frame attached to the rotating mantle (c, d), the Cassini plane is rotating at frequency $-\Omega_c$ in the longitudinal direction.  The unit vector $\mitbf{\hat{e}^p_\perp}(t)$ captures the time-dependent longitudinal orientation of the Cassini plane as seen in the mantle frame; it points in the direction of the projection of $\mitbf{\hat{e}^c_1}$ on the equatorial plane of the mantle.}
\end{center}
\end{figure}

When viewed in the Cassini frame, the orientation of the orbit normal ($\mitbf{\hat{e}^I_3}$) and mantle figure axis ($\mitbf{\hat{e}^p_3}$) remain at fixed orientations with respect to the ecliptic normal ($\mitbf{\hat{e}^c_3}=\mitbf{\hat{e}_3}$) (Fig.~\ref{fig:rotframes}a).  As defined earlier, the angle of tilt between $\mitbf{\hat{e}^p_3}$ and $\mitbf{\hat{e}^c_3}$ is denoted by $\theta_p$ and LLR observations suggest that it is equal to $\theta_p=1.543^\circ$.  This tilt is caused by the gravitational torque that the Earth exerts on the ellipsoidal shape of the mantle and, secondarily, by internal torques from the inner core and the fluid core.  It is the tilt angle that allows to balance the total torque acting on the mantle with a change in its angular momentum at the same rate as the precession of the orbit, and therefore to maintain a stationary configuration in the Cassini frame.

Likewise, the ellipsoidal inner core is also subject to a gravitational torque from Earth and to internal torques from the mantle and fluid core.  For the inner core, the internal torque -- especially the gravitational torque from the fluid core and mantle -- is much more important than the torque from Earth (see DW16).  As is the case for the mantle, the orientation of the figure axis of the inner core (denoted by $\mitbf{\hat{e}^s_3}$) should evolve to that which allows to balance the torque acting on it with a change in its angular momentum at the same rate as the orbit precession. In other words, the inner core is also in a Cassini state and, viewed in the Cassini frame, the orientation of $\mitbf{\hat{e}^s_3}$ remains fixed (Fig.~\ref{fig:rotframes}a).  We expect $\mitbf{\hat{e}^s_3}$ to differ from $\mitbf{\hat{e}^p_3}$ because the inner core is subject to a different torque balance than the mantle.  We define the angle of inner core tilt $\theta_n$ as the angle of misalignment of $\mitbf{\hat{e}^s_3}$ with respect to the mantle figure axis $\mitbf{\hat{e}^p_3}$.

The rotation and symmetry axes of the mantle -- and similarly those of the inner core -- are expected to remain in close alignment, but they do not coincide exactly.  The rotation vector of the fluid core is expected to be misaligned from that of the mantle, remaining instead in a close alignment with the ecliptic normal.  Each of these rotation vectors lie on the Cassini plane and their orientations remain fixed when viewed in the Cassini frame (Fig.~\ref{fig:rotframes}b). We define the rotation vector of the mantle as $\mitbf{\Omega}$, misaligned by an angle $\theta_m$ with respect to the mantle figure axis.  The rotation vectors of the fluid core and inner core are defined as $\mitbf{\Omega_f}$ and $\mitbf{\Omega_s}$.  Their misalignment angles, respectively $\theta_f$ and $\theta_s$, are defined with respect to the mantle rotation vector $\mitbf{\Omega}$ (Fig.~\ref{fig:rotframes}b). 

To be formal in our definition of the different angles of misalignment, $I$ is defined positive pointing from $\mitbf{\hat{e}_3^c}$ to $\mitbf{\hat{e}_3^I}$. Angles $\theta_p$, $\theta_n$, $\theta_m$, $\theta_f$ and $\theta_s$ are defined positive in the clockwise direction when viewed in the Cassini frame.  According to this convention, $\theta_f$ as depicted in Fig.~\ref{fig:rotframes} is negative, and we expect this to be the case since $\mitbf{\Omega_f}$ should be closely aligned with the ecliptic normal ($\mitbf{\hat{e}_3^c}=\mitbf{\hat{e}_3}$).

The mean gravitational torque that Earth exerts on the mantle, averaged over one orbit, can be replaced by that produced by a ring of mass equivalent to that of Earth encircling the Moon on a plane with normal vector $\mitbf{\hat{e}^I_3}$.  Viewed in the Cassini frame, the amplitude of this mean torque remains constant and in direction $-\mitbf{\hat{e}^c_2}$ (the direction of the line of the ascending node), perpendicular to the Cassini plane.  Likewise, the gravitational torque that Earth exerts on the inner core is also perpendicular to the Cassini plane. The direction of the torque depends on the sign of the sum of $(I+\theta_p+\theta_n)$: if it is positive, the torque is in direction $-\mitbf{\hat{e}^c_2}$; if it is negative, the torque is instead in direction $\mitbf{\hat{e}^c_2}$.

Although the mantle figure axis $\mitbf{\hat{e}^p_3}$ remains at a fixed orientation in the Cassini frame, the two equatorial directions $\mitbf{\hat{e}^p_1}$ and $\mitbf{\hat{e}^p_2}$ do not since the mantle is rotating about $\mitbf{\hat{e}^p_3}$.  Viewed in the Cassini frame, the period of rotation of $\mitbf{\hat{e}^p_1}$ and $\mitbf{\hat{e}^p_2}$ around $\mitbf{\hat{e}^p_3}$, must be equal to the time it takes for the Moon to return to the ascending node of its orbit.   The frequency of this rotation, which we denote $\Omega_c$, is equal to $2\pi/27.212$ day$^{-1}$.

We develop our rotational model in a frame attached to the rotating mantle.  As seen by an observer on the mantle, the longitudinal orientation of the Cassini plane is rotating in the retrograde direction about $\mitbf{\hat{e}^p_3}$ at frequency $\Omega_c$ (Fig.~\ref{fig:rotframes}c,d).  The unit vectors $\mitbf{\hat{e}^I_3}$, $\mitbf{\hat{e}^c_3}$ and $\mitbf{\hat{e}^s_3}$ and the rotation vectors $\mitbf{\Omega}$, $\mitbf{\Omega_f}$ and $\mitbf{\Omega_s}$ remain at fixed orientations, but are precessing about $\mitbf{\hat{e}^p_3}$ in the retrograde direction at frequency $\Omega_c$.  Since the gravitational torque by Earth remains perpendicular to the Cassini plane, as seen by an observer on the mantle, this torque is periodic, with a retrograde frequency equal to $\Omega_c$. Following the nutation model of \cite{mathews91a}, it is convenient to introduce a frequency factor $\omega$, connected to $\Omega_c$ by $\Omega_c = -\omega \Omega_o$, where $\Omega_o = 2\pi/27.322$ day$^{-1}$ is the amplitude of the rotation vector of the mantle.  The exact definition of $\Omega_o$ is given by Eq. (\ref{eq:om0a1}) in Appendix A; to a good approximation, $\Omega_o$ is related to $\Omega_c$ and $\Omega_p$ by

\begin{equation}
\Omega_o = \Omega_c - \Omega_p \cos(\theta_p) \, . \label{eq:omega0}
\end{equation}
The frequency factor $\omega$ is then equal to

\begin{equation}
\omega = -\frac{\Omega_c}{\Omega_o} = -1 - \cos(\theta_p)\, \delta \omega \, ,
\label{eq:omega}
\end{equation}
where $\delta \omega = \Omega_p/\Omega_o $= 27.322 days / 18.6 yr = $4.022 \times 10^{-3}$ is the Poincar\'e number, expressing the ratio of precession to rotation frequency.  $\omega$ represents then the frequency of the periodic gravitational forcing that Earth applies on the Moon, expressed in units of cycles per Lunar day, as seen by an observer on the mantle.  

As developed in Appendix A,  the time-dependent longitudinal orientation of the Cassini plane, as seen by an observer on the mantle, and expressed in terms of $\omega$, can be written as 

 \begin{equation}
 \mitbf{\hat{e}^p_\perp}(t) =  \cos (\omega \Omega_o t) \mitbf{\hat{e}^p_1} + \sin (\omega \Omega_o t) \mitbf{\hat{e}^p_2} \, , \label{eq:eperp}
\end{equation}
where $t$ is time and direction $\mitbf{\hat{e}^p_1}$ has been chosen to be aligned with the projection of $\mitbf{\hat{e}^c_1}$ onto the equator of the mantle at $t=0$ (Fig.~\ref{fig:rotframes}c,d).  It can be shown that (see Eq. \ref{eq:dteperpapp})

\begin{equation}
\frac{d}{dt} \mitbf{\hat{e}^p_\perp} (t) =  \omega \Omega_o \Big( \mitbf{\hat{e}^p_3} \times \mitbf{\hat{e}^p_\perp} (t) \Big) \, , \label{eq:dteperp2}
\end{equation}
where the time derivative is taken in the mantle frame. Note that the direction of $\mitbf{\hat{e}^p_3} \times \mitbf{\hat{e}^p_\perp} (t) = \mitbf{\hat{e}^c_2}$ (see Fig.~\ref{fig:rotframes}c,d).  Since  $\omega$ is negative, the time derivative of $\mitbf{\hat{e}^p_\perp} (t)$ points in direction $-\mitbf{\hat{e}^c_2}$, the same direction as the gravitational torque from Earth on the mantle.

Using the definition of $\mitbf{\hat{e}^p_\perp}(t)$ in Eq. (\ref{eq:eperp}), we can express the direction of the normal to the ecliptic $\mitbf{\hat{e}_3}$ and the figure axis of the inner core $\mitbf{\hat{e}^s_3}$, as seen in the mantle frame, by

\begin{subequations}
\begin{align}
\mitbf{\hat{e}_3} = \mitbf{\hat{e}^c_3}  & = \cos \theta_p \,  \mitbf{\hat{e}^p_3} -  \sin \theta_p \, \mitbf{\hat{e}^p_\perp}(t)  \, ,\\
\mitbf{\hat{e}^s_3} & = \cos \theta_n \,  \mitbf{\hat{e}^p_3} +  \sin \theta_n \, \mitbf{\hat{e}^p_\perp}(t) \, .
\end{align}
\label{eq:edef}
\end{subequations}

In appendix A, we present the definitions of the rotation vectors of the mantle, fluid core and inner core for the tidally locked spin-orbit configuration of the Moon.  The amplitude of each of these rotation vectors differ from one another, although the difference between them is small (at most of the order of $\delta \omega$) and it is convenient for the development of our model to approximate all three vectors as having the amplitude $\Omega_o$ defined in Eq. (\ref{eq:omega0}).  The rotation vectors of the mantle, fluid core and inner core can be written, respectively, as
           
\begin{subequations}
\begin{align}
\mitbf{\Omega} & = \Omega_o \bigg( \cos \theta_m \,  \mitbf{\hat{e}^p_3} +  \sin \theta_m \, \mitbf{\hat{e}^p_\perp}(t) \bigg) \, ,\\
\mitbf{\Omega_f} & = \Omega_o \bigg( \cos (\theta_m+\theta_f) \mitbf{\hat{e}^p_3} +  \sin (\theta_m+\theta_f)  \mitbf{\hat{e}^p_\perp}(t) \bigg) \, ,\\
\mitbf{\Omega_s} & =\Omega_o \bigg( \cos (\theta_m+\theta_s) \mitbf{\hat{e}^p_3} +  \sin (\theta_m+\theta_s)  \mitbf{\hat{e}^p_\perp}(t) \bigg) \, .
\end{align}
\label{eq:omdef}
\end{subequations}
It is further convenient to introduce ${\mitbf \omega_f}$ and ${\mitbf \omega_s}$, the perturbation in the rotation of the fluid core and inner core, respectively, with respect to that of the mantle, defined as

\begin{subequations}
\begin{align}
{\mitbf \omega_f} &=  {\mitbf \Omega_f}  - {\mitbf \Omega}  \, ,\label{eq:omf} \\
{\mitbf \omega_s} &=  {\mitbf \Omega_s}  - {\mitbf \Omega} \, .\label{eq:oms} 
\end{align}
\label{eq:omegafs}
\end{subequations}

\subsection{The rotational model}

Our goal is to determine the Cassini state of the whole of the Moon.  That is, to determine the precession dynamics of our reference interior model of the Moon when placed in orbit around Earth and subject to its gravitational torque.  In short, our goal is to determine the five angles $\theta_p$, $\theta_n$, $\theta_m$, $\theta_f$ and  $\theta_s$ for a given Lunar interior density structure. These angles obey a system of five equations.  The first three describe respectively the evolution of the angular momentum of the whole Moon ($\mitbf{H}$), the fluid outer core ($\mitbf{H_f}$) and solid inner core ($\mitbf{H_s}$) in the reference frame rotating with the mantle, 

\begin{subequations}
\begin{align}
 & \frac{d}{dt} \mitbf{H}  +  \mitbf{\Omega} \times \mitbf{H} = \mitbf{\Gamma} \, ,
\label{eq:am} \\
 & \frac{d}{dt} \mitbf{H_f}  -  \mitbf{\omega_f} \times \mitbf{H_f}
=  \mitbf{0}\, ,
\label{eq:amfoc}  \\
 & \frac{d}{dt} \mitbf{H_s}  +  \mitbf{\Omega} \times \mitbf{H_s} =
\mitbf{\Gamma_s} \, ,
\label{eq:amsic}
\end{align}
\label{eq:amall}
\end{subequations}
where $\mitbf{\Gamma}$ is the gravitational torque from Earth acting on the whole Moon and $\mitbf{\Gamma_s}$ is the total gravitational and pressure torque exerted on the inner core.  The final two equations of the model are kinematic relations, one to express the change in the orientation of the inner core figure resulting from its own differential rotation, and the second describing the invariance of the ecliptic normal in the inertial frame as seen in the frame attached to the mantle.  They are respectively, 

\begin{subequations}
\begin{align} 
& \frac{d}{dt} \mitbf{\hat{e}^s_3} + \mitbf{\hat{e}^s_3} \times  \mitbf{\omega_s}  = \mitbf{0} \, , \label{eq:kins}\\
& \frac{d}{dt} \mitbf{\hat{e}_3} + \mitbf{\Omega} \times  \mitbf{\hat{e}_3}  = \mitbf{0} \, .\label{eq:kinp} 
\end{align}
\label{eq:kinall}
\end{subequations}

The combination of Eqs. (\ref{eq:amall}) and (\ref{eq:kins}) forms the foundation of the nutation model of \cite{mathews91a} that takes into account internal coupling between inner core, fluid core and mantle subject to an external torque.  Eq. (\ref{eq:kinp}) allows us to connect this model to the tilt of the figure axis of the Moon's mantle to the ecliptic.  Note that Eq. (\ref{eq:kins}) is different from the one used in DW16; we use here the original equation of the nutation model of \cite{mathews91a} (see their equation 19).  Also note that in DW16, Eq. (\ref{eq:kinp}) was replaced by a second dynamical equation for the Moon, similar to Eq. (\ref{eq:am}), but viewed in the ecliptic frame. It was shown that the two dynamical equations for the Moon were tied by the condition expressed in Eq. (\ref{eq:kinp}), introduced by \cite{eckhardt81}, and it is more convenient to simply use the latter here.

Definitions for $\mitbf{H}$, $\mitbf{H_f}$ and $\mitbf{H_s}$ are given by

\begin{subequations}
\begin{align}
{\mitbf H} &=  \mitbf{\cal I} \cdot {\mitbf \Omega} +  \mitbf{{\cal I}_f} \cdot {\mitbf \omega_f} +  \mitbf{{\cal I}_s} \cdot {\mitbf \omega_s} \, ,\label{eq:H}\\
{\mitbf H_f} &=  \mitbf{{\cal I}_f} \cdot {\mitbf \Omega_f} \, ,\label{eq:Hf} \\
{\mitbf H_s} &=  \mitbf{{\cal I}_s} \cdot {\mitbf \Omega_s} \, ,\label{eq:Hs}
\end{align}
\label{eq:allH}
\end{subequations}
where $\mitbf{{\cal I}_s}$, $\mitbf{{\cal I}_f}$, and $\mitbf{\cal I}$ are the moment of inertia tensors of the solid inner core, fluid core and the whole Moon, respectively.   Explicit definitions for these are given in Eq. A9 of DW16; they involve the principal moments of inertia of the whole Moon, fluid core and solid inner core. 

We neglect the triaxial shape of the Moon in the development of the expression of the angular momentum vectors of each region.  In other words, we assume that the two equatorial moments of inertia are equal to one another and given by the mean values defined in Eq. (\ref{ref:meanH}). We also neglect elastic deformations.  Proceeding this way, the expansion of the angular momentum vectors gives 
 
\begin{subequations}
\begin{align}
 \mitbf{H_s} & = \bar{A}_s \Omega_o \Big[ \cos(\theta_m + \theta_s) \mitbf{\hat{e}^p_3} + \sin (\theta_m + \theta_s)\mitbf{\hat{e}^p_\perp}(t) \Big] \, \nonumber\\
 &+ \bar{A}_s e_s \Omega_o \cos(\theta_n - \theta_m - \theta_s) \Big[ \cos(\theta_n) \mitbf{\hat{e}^p_3} + \sin (\theta_n)\mitbf{\hat{e}^p_\perp}(t) \Big] \, , \label{eq:Hs2} \\
 \mitbf{H_f} & = \bar{A}_f \Omega_o \Big[ (1+e_f) \cos(\theta_m + \theta_f) \mitbf{\hat{e}^p_3} + \sin (\theta_m + \theta_f)\mitbf{\hat{e}^p_\perp}(t) \Big] \, \nonumber\\
 &- \alpha_1 \bar{A}_s e_s \Omega_o \cos(\theta_n - \theta_m - \theta_f) \Big[ \cos(\theta_n) \mitbf{\hat{e}^p_3} + \sin (\theta_n)\mitbf{\hat{e}^p_\perp}(t) \Big] \, , \nonumber\\
 &+ \alpha_1 \bar{A}_s e_s \Omega_o \cos(\theta_m + \theta_f)  \mitbf{\hat{e}^p_3}  \,  \label{eq:Hf2} \\
\mitbf{H} & = \Omega_o \Big[ (C -C_f -C_s) \cos (\theta_m) \mitbf{\hat{e}^p_3}+ (\bar{A} -\bar{A}_f -\bar{A}_s) \sin (\theta_m) \mitbf{\hat{e}^p_\perp}(t) \Big] + \mitbf{H_f} +\mitbf{H_s} \, ,
 \label{eq:H2} 
\end{align}
\label{eq:Hall2}
\end{subequations}
where $\alpha_1$ is related to the density contrast between the solid and fluid core. The coefficient $\alpha_1$ and the related coefficient $\alpha_3 = 1- \alpha_1$ that we introduce below are defined in Eq. A8 of DW16. For uniform density layers, they simplify to 

\begin{equation}
\alpha_1 = \frac{\rho_f}{\rho_s} \, , \hspace*{1cm} \alpha_3  =  1 - \frac{\rho_f}{\rho_s} \, .
\label{eq:alpha13}
\end{equation}

As explained in the previous section, the gravitational torque from Earth points in direction $- \mitbf{\hat{e}^p_3} \times \mitbf{\hat{e}^p_\perp}(t) $. Hence, we can write the gravitational torque acting on the whole of the Moon as

\begin{subequations}
\begin{equation}
\mitbf{\Gamma}  = - \Gamma \, \Big( \mitbf{\hat{e}^p_3} \times \mitbf{\hat{e}^p_\perp}(t) \Big) \, ,
\end{equation}
where $\Gamma$ is the amplitude of the torque averaged over one orbit. Valid to second order in ellipticity, it is equal to

\begin{align}
{\Gamma}  & =   \frac{3}{2} \frac{ {\cal M} n^2}{(1-e_L^2)^{3/2}} \bigg[ \big( C-A \big) - \big(C_s -A_s \big)\alpha_3 \bigg]\sin(I + \theta_p)\cos(I + \theta_p) \, \nonumber\\
& +  \frac{3}{2} \frac{{\cal M} n^2}{(1-e_L^2)^{3/2}} \bigg[ \big( C_s-A_s \big) \alpha_3 \bigg]\sin(I + \theta_p + \theta_n)\cos(I + \theta_p + \theta_n) \, \nonumber\\
& +  \frac{3}{8} {\cal M} n^2 \bigg[ \big( B - A \big) - \big( B_s - A_s \big) \alpha_3 \bigg] \left( 1 - \frac{5}{2} e_L^2 - \Big(1 + \frac{11}{2} e_L^2 \Big) \cos(I + \theta_p) \right) \sin(I + \theta_p)  \nonumber\\
& + \frac{3}{8} {\cal M}  n^2 \bigg[ \big( B_s - A_s \big)\alpha_3 \bigg] \left( 1 - \frac{5}{2} e_L^2 - \Big(1 + \frac{11}{2} e_L^2 \Big) \cos(I + \theta_p + \theta_n) \right) \sin(I + \theta_p +\theta_n) \, , \label{eq:tqearth}
\end{align}
\label{eq:tqearthboth}
\end{subequations}
where $e_{L}$ is the orbit eccentricity, $n$ is the mean motion of the Moon, and ${\cal M} = M_E/ (M+M_E)$, where $M_E$ is the mass of Earth.  In the absence of an inner core ($C_s=B_s=A_s=0$), the torque in Eq. (\ref{eq:tqearth}) is equal to that given in \cite{peale69}. Because of the synchronous rotation of the Moon around Earth, the torque involves the full triaxial definition of the moment of inertia.  For small $(I + \theta_p)$, the last two terms of Eq. (\ref{eq:tqearth}) are small compared to the first two terms, and they were neglected in DW16.  

Likewise, the torque acting on the inner core can be written as

\begin{subequations}
\begin{equation}
\mitbf{\Gamma_s}  = - \Gamma_s \, \Big( \mitbf{\hat{e}^p_3} \times \mitbf{\hat{e}^p_\perp}(t) \Big) \, .
\end{equation}
Valid to second order in ellipticity, the amplitude of the torque $\Gamma_s$ is 

\begin{align}
{\Gamma_s} & =    \frac{3}{2} \frac{ {\cal M} n^2}{(1-e_L^2)^{3/2}} \big( C_s-A_s \big) \alpha_3 \, \sin(I + \theta_p + \theta_n)\cos(I + \theta_p + \theta_n) \nonumber\\
& + \frac{3}{8} {\cal M}n^2   \big( B_s - A_s \big) \alpha_3 \left( 1 - \frac{5}{2} e_L^2 - \Big(1 + \frac{11}{2} e_L^2 \Big) \cos(I + \theta_p +\theta_n) \right) \sin(I + \theta_p+\theta_n)  \nonumber\\
&  + \Omega_o^2  \bar{A}_s e_s  \alpha_3 \alpha_g  \sin(\theta_n)\cos(\theta_n)  \nonumber\\
&   + \Omega_o^2 \bar{A}_s  e_s \alpha_1 \, \sin(\theta_m + \theta_f - \theta_n)\cos(\theta_m + \theta_f - \theta_n ) \,  , \label{eq:tqs}
\end{align}
\label{eq:tqsboth}
\end{subequations}
where the coefficient $\alpha_g$ captures the strength of gravitational coupling by the rest of the Moon on a tilted inner core.  This coefficient is derived in \cite{mathews91a}, and is also defined in Eq. A14b of DW16; for uniform density layers, it simplifies to 

\begin{equation}
\alpha_g  = \frac{8\pi G}{5\Omega_o^2} \left[ \rho_c (\epsilon_r - \epsilon_m) + \rho_m (\epsilon_m - \epsilon_f) + \rho_f \epsilon_f \right] \, ,
\label{eq:alphag}
\end{equation}
where $G$ is the gravitational constant.  The first two terms that enter Eq. (\ref{eq:tqs}) represent the gravitational torque from Earth.   The last two represent, respectively, the gravitational torque from the mantle and fluid core and the pressure torque at the inner-core boundary.  In contrast to the torque from Earth, these internal torques involve the mean equatorial moment of inertia.  This is because these torques result from the precession between the different layers.  Thus, over one orbit, they involve an average of the torque about $A_s$ and $B_s$.  

Using the definition of the torques in Eqs. (\ref{eq:tqearthboth}) and (\ref{eq:tqsboth}), the three angular momentum equations of Eqs. (\ref{eq:amall}) and the two kinematic relations of Eqs. (\ref{eq:kinall}) form the following set of five conditions, 

\begin{subequations}
\begin{align}
&\bar{A} \Bigl[ \Bigl( \omega - e \cos(\theta_m) \Bigr) \sin(\theta_m) \Bigr]  \nonumber\\
      + \, &\bar{A}_f \Bigl[\sin(\theta_f) +  \omega \Bigl( \sin(\theta_m +\theta_f) - \sin(\theta_m) \Bigr) -e_f \sin(\theta_m) \Bigl( \cos(\theta_m +\theta_f) - \cos(\theta_m) \Bigr) \Bigr]  \nonumber\\
      + \,& \bar{A}_s \Bigl[  \sin(\theta_s) +  \omega \Bigl( \sin(\theta_m +\theta_s) - \sin(\theta_m) \Bigr) -e_s \sin(\theta_m) \Bigl( \alpha_1 \cos(\theta_m +\theta_f) - \cos(\theta_m) \Bigr) \Bigr]  \nonumber\\
      + \,&  \bar{A}_s   e_s \alpha_3 \cos(\theta_n - \theta_m - \theta_f) \Bigl( \omega \sin(\theta_n)  +  \sin(\theta_n-\theta_m) \Bigr) \nonumber\\
       = \,&- \Phi_\beta^p  \Bigl( \bar{A} \beta -  \bar{A}_s   \beta_s \alpha_3 \Bigr) - \Phi_\beta^n   \bar{A}_s   \beta_s \alpha_3 - \Phi_\gamma^p  \Bigl( \bar{A} \gamma -  \bar{A}_s  \gamma_s \alpha_3 \Bigr) - \Phi_\gamma^n   \bar{A}_s  \gamma_s \alpha_3 \, , \label{eq:f1}
\end{align}

\begin{align}
& \bar{A}_f \Bigl[ \sin(\theta_f) +  \omega  \sin(\theta_m +\theta_f) + e_f \cos(\theta_m +\theta_f) \Bigl( \sin(\theta_m +\theta_f) - \sin(\theta_m) \Bigr) \Bigr]  \nonumber\\
+ \, &   \bar{A}_s e_s \alpha_1 \Bigl[  \cos(\theta_n -\theta_m -\theta_f) \Bigl( - \omega  \sin(\theta_n) -\sin(\theta_n -\theta_m) - \sin(\theta_m+\theta_f-\theta_n) \Bigr) \Bigr]  \nonumber\\
+ \, &   \bar{A}_s e_s \alpha_1 \Bigl[  \cos(\theta_m +\theta_f) \Bigl( \sin(\theta_m +\theta_f) - \sin(\theta_m) \Bigr) \Bigr] = 0 \, , \label{eq:f2}
\end{align}

\begin{align}
&  \Big[ \sin(\theta_s) +  \omega  \sin(\theta_m +\theta_s) + e_s \alpha_3 \alpha_g \sin(\theta_n) \cos(\theta_n) \Big]  \nonumber\\
+ \, & e_s\cos(\theta_n -\theta_m -\theta_s) \Big[ \omega  \sin(\theta_n) +\sin(\theta_n -\theta_m) \Big] \nonumber\\
- \, & e_s  \cos(\theta_n -\theta_m -\theta_f) \Big[  \alpha_1 \sin(\theta_n-\theta_m-\theta_f) \Big] \nonumber\\
= \, & - \Phi_\beta^n   \beta_s \alpha_3 - \Phi_\gamma^n   \gamma_s \alpha_3  \, , \label{eq:f3}
\end{align}

\begin{equation}
 \omega  \sin(\theta_n) + \sin(\theta_m +\theta_s - \theta_n) - \sin(\theta_m -\theta_n) = 0 \, , \label{eq:f4}
\end{equation}

\begin{equation}
\omega  \sin(\theta_p) +  \sin(\theta_m +\theta_p) =0 \, , \label{eq:f5}
\end{equation}
\label{eq:condall}
\end{subequations}
where we have defined 

\begin{subequations}
\begin{align}
\Phi_\beta^p  & = \frac{3}{2}  \frac{{\cal M}}{(1-e_L^2)^{3/2}} \sin(I + \theta_p)\cos(I + \theta_p) \, , \\
\Phi_\beta^n  & = \frac{3}{2}  \frac{{\cal M}}{(1-e_L^2)^{3/2}} \sin(I + \theta_p + \theta_n)\cos(I + \theta_p+\theta_n) \, , \\
\Phi_\gamma^p  & = \frac{3}{8} {\cal M}\left( 1 - \frac{5}{2} e_L^2 - \Big(1 + \frac{11}{2} e_L^2 \Big) \cos(I + \theta_p) \right) \sin(I + \theta_p) \, , \\
\Phi_\gamma^n  & = \frac{3}{8} {\cal M} \left( 1 - \frac{5}{2} e_L^2 - \Big(1 + \frac{11}{2} e_L^2 \Big) \cos(I + \theta_p + \theta_n) \right) \sin(I + \theta_p + \theta_n) \, ,
\end{align}
\label{eq:phis}
\end{subequations}

and

\begin{subequations}
\begin{align}
& \beta = \frac{C-A}{B} \approx \frac{C-A}{\bar{A}} \, , \hspace*{1cm}  
\beta_s = \frac{C_s-A_s}{B_s} \approx \frac{C_s-A_s}{\bar{A}_s} \, , \\
& \gamma = \frac{B-A}{B} \approx \frac{B-A}{\bar{A}} \, , \hspace*{1cm}  
\gamma_s = \frac{B_s-A_s}{B_s} \approx \frac{B_s-A_s}{\bar{A}_s} \, .
\end{align}
\label{eq:betagamma}
\end{subequations} 
Note that the mantle rotation rate $\Omega_o$ is approximately equal to the sidereal frequency $n$ and we have set $n=\Omega_o$, which removes a factor of $n^2 / \Omega_o^2$ multiplying the right-hand sides of Eqs. (\ref{eq:f1}) and (\ref{eq:f3}). The five conditions of Eqs. (\ref{eq:condall}) constitute the set of non-linear conditions on the five angles $\theta_p$, $\theta_n$, $\theta_m$, $\theta_f$ and  $\theta_s$ that must be simultaneously satisfied to determine the complete Cassini state of the Moon.   In the limit of small angles, 

\begin{equation}
\cos (\theta_i) \rightarrow 1 \, , \hspace*{1cm} \sin (\theta_i) \rightarrow \theta_i \, ,
\end{equation}
and for $\Phi_\gamma^p = \Phi_\gamma^n = 0$, we retrieve the linear system of equations presented in DW16, where the parameter ${\cal M}$ was omitted, and where the parameter $\beta_s$ that appears in Eqs. (\ref{eq:f1}) and (\ref{eq:f3}) was approximated as $e_s$.  

For a Moon model with no core, the system of conditions reduces to

\begin{subequations}
\begin{align} 
& \bar{A}  \Bigl( \omega - e \cos(\theta_m) \Bigr) \sin(\theta_m) = - \Phi_\beta^p  \bar{A} \beta   - \Phi_\gamma^p  \bar{A} \gamma  \, , \\
& \omega  \sin(\theta_p) +  \sin(\theta_m +\theta_p) = 0 \, ,
\end{align}
\label{eq:condnocore}
which can be combined to form 

\begin{equation}
\bar{A} \Big(\omega - e \cos (\theta_m) \Big) \Big(-\omega - \cos (\theta_m) \Big) \tan (\theta_p) =  - \Phi_\beta^p  \bar{A} \beta   - \Phi_\gamma^p  \bar{A} \gamma  \, .
\end{equation}
Using $C=\bar{A}(1+e)$, $\omega$ defined in Eq. (\ref{eq:omega}), and also that $\Omega_p / \Omega_o  \ll 1$ and $\theta_m \ll 1$, we retrieve (in our notation) the condition on $\theta_p$ given in Eq. (19) of \cite{peale69} that defines the Cassini state of a single body Moon 

\begin{equation}
{C} \frac{\Omega_p}{\Omega_o} \sin (\theta_p) =   \Phi_\beta^p  \bar{A} \beta  +  \Phi_\gamma^p  \bar{A} \gamma  \, . \label{eq:cass}
\end{equation}  
\end{subequations} 
We show in Appendix B an alternate derivation of Eq. (\ref{eq:cass}), one based on considering the rotational dynamics in the inertial frame.

A condition similar to Eq. (\ref{eq:cass}) but for the inner core of the Moon can be derived. Before we do this, it is convenient to introduce here the frequency of the FICN, $\omega_{ficn}$, which as we show below, turns out to be a fundamental component of the Cassini state of the inner core.  The FICN describes the free precession of the spin-symmetry axis of the inner core when it is misaligned from the mantle.  The FICN frequency depends on the sum of the torques exerted on the inner core and, when expressed in cycles per Lunar day, it is approximately equal to (see DW16) 

\begin{equation}
\omega_{ficn}  = e_s \alpha_1 - e_s \alpha_g \alpha_3 - \frac{3}{2} \frac{\beta_s \alpha_3}{(1-e_L^2)^{3/2}} (\cos^2 I - \sin^2 I)  \, . \label{eq:omficn}
\end{equation}
For the Moon, the gravitational torque exerted by the fluid core and mantle on the inner core (second term on the right-hand side of Eq. \ref{eq:omficn}) is much larger than the pressure torque at the ICB and the gravitational torque from Earth (first and third terms of Eq. \ref{eq:omficn}, respectively), so $\omega_{ficn}$ is negative and the FICN mode is retrograde. 

It is also convenient to derive alternate forms of conditions  (\ref{eq:f4}) and (\ref{eq:f5}).  First, using the definition of $\omega$ in Eq. (\ref{eq:omega}) and $\cos({\theta_m}) \rightarrow 1$ allows one to write the condition of Eqs. (\ref{eq:f5}) as  

\begin{subequations}
\begin{equation}
\sin (\theta_m) = \delta \omega \, \sin (\theta_p) \, .\label{eq:sinm}
\end{equation}
This expresses the connection between the misalignment of the rotation vector of the mantle from its figure axis and the tilt of the latter with respect to the ecliptic normal.  They are related by the Poincar\'e number $\delta \omega$.  Because the Poincar\'e number is small, $\theta_m \ll\theta_p$.  Using this, the condition of Eq.~(\ref{eq:f4}) can be written as 

\begin{equation}
\sin (\theta_m+\theta_s-\theta_n) = \delta \omega \, \sin (\theta_p+\theta_n) \, ,\label{eq:sinmsn}
\end{equation}
which is the analogous relationship for the inner core, connecting in the same manner the angle of misalignment of its rotation vector from its figure axis ($\theta_m+\theta_s-\theta_n$) to the tilt of its figure axis with respect to the ecliptic normal.

The Cassini state of the inner core can be derived on the basis of its angular momentum balance (Eq.~\ref{eq:f3}). 
Using Eqs. (\ref{eq:sinm}-\ref{eq:sinmsn}), and setting $\theta_s\approx \theta_n$ (see DW16), one can show that 

\begin{align}
& \sin(\theta_s) + \omega \sin(\theta_s + \theta_m) \approx - \delta \omega \sin(\theta_p + \theta_n) \, , \\
& e_s \cos(\theta_n-\theta_m-\theta_s) \left[ \omega \sin(\theta_n) + \sin(\theta_n-\theta_m) \right] \approx - e_s \delta \omega \sin(\theta_p + \theta_n) \, ,
\end{align}
\end{subequations}
so that Eq.~(\ref{eq:f3}) can be written as

\begin{subequations}
\begin{align}
 &- (1+e_s)\,  \delta \omega \, \sin(\theta_p + \theta_n)\, + \, e_s \alpha_3 \alpha_g \sin(\theta_n) \cos(\theta_n) \, \nonumber\\
 & - \, e_s \alpha_1 \cos(\theta_n -\theta_m -\theta_f) \sin(\theta_n-\theta_m-\theta_f)  = \,  - \Phi_\beta^n   \beta_s \alpha_3 - \Phi_\gamma^n   \gamma_s \alpha_3  \, . \label{eq:f3alt}
\end{align}
On using $\delta \omega = \Omega_p/\Omega_o$, $C_s=\bar{A}_s (1+e_s)$,  $\theta_m+\theta_f\approx-\theta_p$ (expressing the fact that the rotation vector of the fluid core remains almost aligned with the ecliptic normal), Eq. (\ref{eq:f3alt}) becomes

 \begin{align}
 \frac{C_s}{\bar{A}_s} & \frac{\Omega_p}{\Omega_o} \,  \sin (\theta_p+\theta_n) = \nonumber\\
&   \Phi_\beta^n  \beta_s \alpha_3  \, + \, \Phi_\gamma^n \gamma_s \alpha_3 \, + \, e_s  \alpha_3 \alpha_g  \sin(\theta_n)\cos(\theta_n) - \, e_s \alpha_1 \, \sin(\theta_n + \theta_p)\cos(\theta_n + \theta_p ) \, .\label{eq:casssic1}
 \end{align}
This last equation determines the Cassini state of the inner core of the Moon.  In Appendix B, we show how the same condition can be derived, perhaps more simply, by considering the dynamics in the inertial reference frame.  Because internal torques dominate the gravitational torque from Earth in the present-day Moon (DW16), Eq.~(\ref{eq:casssic1}) can be further simplified if we set $\Phi_\beta^n  = \Phi_\gamma^n = 0$.  As our results will confirm, $\theta_n$ is typically much larger than $\theta_p=1.543^\circ$, so we can approximate $\sin(\theta_n + \theta_p)\cos(\theta_n + \theta_p )$ as $\sin(\theta_n)\cos(\theta_n)$.  Furthermore, since the dynamical ellipticity of the inner core is small, $C_s \approx \bar{A}_s$. Upon using the expression of the FICN frequency $\omega_{ficn}$ given by Eq. (\ref{eq:omficn}), the Cassini state of the inner core of the Moon simplifies to 

\begin{equation}
\frac{\Omega_p}{\Omega_o}  \, \sin (\theta_p+\theta_n) + \omega_{ficn} \sin(\theta_n)\cos(\theta_n) \,  =0 \, . \label{eq:ficncond}
 \end{equation}
\end{subequations}
As we will show, this last equation provides a very good prediction of the tilt angle of the inner core $\theta_n$.  Importantly, it shows that the interior density structure of the Lunar interior influences $\theta_n$ only through the way in which it affects $\omega_{ficn}$; different interior models of the Moon that share the same $\omega_{ficn}$ have the same $\theta_n$. 

Before we present results, a few points about our model are worth noting.  First, we have neglected all elastic deformations in our derivation, assuming that solid regions are perfectly rigid.  The $k_2$ Love number of the Moon is small, approximately $0.02$ \citep{williams14}, thus assuming a rigid mantle is not a bad approximation.  However, elastic (or viscoealstic) deformations deep inside the Moon may be important. 

Second, we have adopted an oversimplified representation of flow motion in the fluid core, restricted to a simple solid body rotation.  In truth, the fluid core can sustain different types of waves, including inertial waves, which can interact with, and alter the FCN and FICN precession modes \cite[e.g.][]{rogister09}. 

Third, although we have retained the triaxial shape of the Moon in the expression of the mean torque from Earth, the angular momentum response is based on axially symmetric model.  The convenience of doing this is that, for each region, we can combine the two equatorial angular momentum equations into a single equation.  To first order, considering the fully triaxial shape of the Moon should not alter much the frequency of the FCN \cite[e.g.][]{vanhoolst02}.  By extension, we assume here that the other free precession mode with a retrograde period close to one Lunar day (when seen in the rotating mantle frame), the FICN, is also not significantly altered by triaxiality. Since the orientations of the fluid core spin axis and the inner core spin-symmetry axis are primarily determined by the FCN and FICN frequencies, respectively, our axially symmetric model should, to first order, capture the salient features of the Cassini state.

\section{Results}

\subsection{Interior Moon models}

The numerical values for the Lunar parameters used in our calculations are listed in Table \ref{table:parameters}.  To compute all other parameters that enter our rotational model, we need to build models of the interior density structure of the Moon.  The first step involves to determine the radial density structure.  We assume a mean Lunar radius of $R=1737.151$ km \cite[][]{williams14}.  We then choose values for the inner core radius ($r_s$), fluid core radius ($r_f$) and crustal thickness ($h_c$) and values for the density of the inner core ($\rho_s$) and crust ($\rho_c$).  The density of the mantle ($\rho_m$) is then determined by matching the moment of inertia of the solid Moon $I_{sm}$.  The value of $I_{sm}$ from \citet[][]{williams14} listed in Table \ref{table:parameters} in principle includes a contribution from the inner core, though it is small compared to that of the outer shell (mantle and crust).  Here, we assume that $I_{sm}$ represents the moment of inertia of the mantle and crust alone and calculate $\rho_m$ using Eq. (13) of DW16.  The density of the fluid core ($\rho_f$) is then found by matching the bulk mass of the Moon $M = (4 \pi/3) \bar{\rho} R^3$, where $\bar{\rho}$ is the mean density, using Eq. (12) of DW16.  Once all radii and densities are defined, the mean equatorial moments of inertia $\bar{A}$, $\bar{A}_f$ and $\bar{A}_s$ are calculated from Eq. (14) of DW16.

\begin{table}
\begin{tabular}{ll}
\hline
Moon Parameter & Numerical value  \\ \hline
rotation rate, $\Omega_o$ &  $2.6617 \times 10^{-6}$ s$^{-1}$ \\
orbit precession rate, $\Omega_p$ & $2\pi /18.6$ yr$^{-1}$ \\
Poincar\'e number, $\delta \omega = {\Omega_p}/{\Omega_o}$ & $4.022 \times 10^{-3}$ \\
mean planetary radius, $R$ & $1737.151$ km\\
mass, $M$ & $7.3463 \times 10^{22}$ kg \\
mean density, $\bar{\rho}$ & $3345.56$ kg m$^{-3}$ \\
moment of inertia of solid Moon, $I_{sm}$ & $0.393112 \cdot M R^2$ \\$J_2$  & $2.03504 \times 10^{-4}$ \\
$C_{22}$ & $2.24482 \times 10^{-5}$ \\
polar surface flattening, $\epsilon_r$ & $1.2899 \times 10^{-3}$\\
equatorial surface flattening, $\xi_r$ & $2.4346 \times 10^{-4}$\\
\end{tabular}
\caption{\label{table:parameters} Reference parameters for the Moon.  The values of $R$, $M$, $\bar{\rho}$, $I_{sm}$, $J_2$ and $C_{22}$ are taken from \cite{williams14}. The values for the unnormalized potential coefficients  $J_2$ and $C_{22}$ include the permanent tide from synchronous rotation with Earth, and are obtained after multiplying the reported values in \cite{williams14} by a factor 1.000978 to take into account our choice of using the mean planetary radius as the reference radius for our calculations instead of the reference radius of $1738$ km used in the GRAIL-derived gravity field. $\epsilon_r$ and $\xi_r$ are taken from \cite{araki09} and converted to our choice of normalization.}
\end{table}

The second step is to determine the polar ($\epsilon$) and equatorial ($\xi$) flattenings at all boundaries.  These are determined on the basis of the reference Moon model defined in section 2.1 in which the principal moments of inertia of each regions are aligned.  We assume that both the ICB and CMB are at hydrostatic equilibrium, in which case their flattenings can be written in terms of the flattenings at the surface and crust-mantle boundary as given by Eqs. (18-20) of DW16.  Under this assumption, the expression for $J_2$ given by Eq. (\ref{eq:j2}) can be written in terms of $\epsilon_r$ and $\epsilon_m$, and likewise, $C_{22}$ given by Eq. (\ref{eq:c22}) can be written in terms of $\xi_r$ and $\xi_m$.   We use the surface flattenings $\epsilon_r = 1.2899 \times 10^{-3}$ and $\xi_r = 2.4346 \times 10^{-4}$ corresponding to the (normalized) topography spherical harmonic coefficients $c_{20}$ and $c_{22}$ taken from \cite[][]{araki09}. The values of $\epsilon_m$ and $\xi_m$ are then determined by matching the observed values of $J_2$ and $C_{22}$ (see Table \ref{table:parameters}).   The values of ($\epsilon_s$, $\xi_s$) and ($\epsilon_f$, $\xi_f$) are then computed from ($\epsilon_r$, $\xi_r$) and ($\epsilon_m$, $\xi_m$) based on the assumption of hydrostatic equilibrium.  Once the polar flattening of each boundary is known, $\alpha_g$ can be determined from Eq. (\ref{eq:alphag}) and the dynamical ellipticities $e_s$, $e_f$ and $e$ defined in Eq. (\ref{eq:e}) are then computed from Eq. (15) of DW16.

The parameters $\beta$ and $\gamma$ defined in Eq. (\ref{eq:betagamma}) that are involved in the torque from Earth are related to $J_2$ and $C_{22}$ by

\begin{equation}
\beta = e \left(1 + 2 \frac{C_{22}}{J_2} \right) \, , \hspace*{1cm} \gamma = 4 e \frac{C_{22}}{J_2} \, .
\end{equation}
The parameters $\beta_s$ and $\gamma_s$ are directly related to the polar and equatorial flattenings at the ICB through

\begin{equation}
\beta_s  = \epsilon_s +\frac{\xi_s}{2}  \, , \hspace*{1cm}
\gamma_s  =  \xi_s \,  . \label{eq:betagammas}
\end{equation}

There is a small inconsistency in our procedure that must be pointed out.  The contribution of the inner core to $J_2$ and $C_{22}$, as written in Eq. (\ref{eq:j2c22}), assumes an inner core aligned with the mantle.  These expressions should really involve the average over one orbit of the polar and equatorial flattenings of a tilted inner core.  However, these depend on the angle of tilt of the inner core, which a-priori we do not know.  This implies that the amplitude of the torque from Earth on the inner core determined by $\beta_s$ and $\gamma_s$ in Eq.~\ref{eq:betagammas} is slightly incorrect.  However, because the torque that the mantle and fluid core exerts on a tilted inner core is much larger than the torque from Earth, this inconsistency has little influence on the results presented in the next section.

\subsection{The Cassini states associated with the inner core}

The set of five conditions in Eqs. (\ref{eq:f1}-\ref{eq:f5}) is solved by a Newton-Raphson method for nonlinear systems \cite[e.g.][]{numrec}.  Each solution presented below is obtained with initial guesses for $\theta_p$, $\theta_m$ and $\theta_f$ taken as $1.5^\circ$,  $0^\circ$ and  $-1.5^\circ$, respectively.  The initial guess for $\theta_n$ is set equal to $\theta_s$ and chosen randomly between $-90^\circ$ and $90^\circ$.  For each set of model parameters, to ensure all possible solutions are found, we repeat the search with a number of random initial guesses for $\theta_n$ (typically 50).  Solutions for which any of the five angles falls outside the bounds of $\left[-90^\circ, 90^\circ\right]$ are discarded.  

We also present results based on a small-angle limit of our model, by taking $\cos(\theta_i)  \approx 1$ and  $\sin(\theta_i)  \approx \theta_i$ for each of the five  angles, and using the following approximations 

\begin{subequations}
\begin{align}
& \sin(I + \theta_p) \cos(I + \theta_p)   \approx \cos I \sin I + \left( \cos^2I - \sin^2I \right) \theta_p \, , \label{eq:app1}\\
& \sin(I + \theta_p + \theta_n) \cos(I + \theta_p +\theta_n)   \approx \cos I \sin I + \left( \cos^2I - \sin^2I \right) \left( \theta_p +\theta_n \right) \, ,\label{eq:app2}\\
& \sin(I + \theta_p) \approx \sin I + ( \cos I ) \theta_p \, . \label{eq:app3}
\end{align}
\label{eq:approxangle}
\end{subequations}
In this small-angle limit, the model is now linear in the five unknown angles.  Note that this small-angle solution is very close, but not exactly equal to that from the model presented in DW16.  The difference is caused by the addition here of the $\gamma$ and $\gamma_s$ terms in the torque from Earth, the inclusion of the factor ${\cal M}$ in the amplitude of the torque, and because in DW16 the parameter $\beta_s$ was approximated as $e_s$.

\begin{figure}
\begin{center}
\includegraphics[height=12cm]{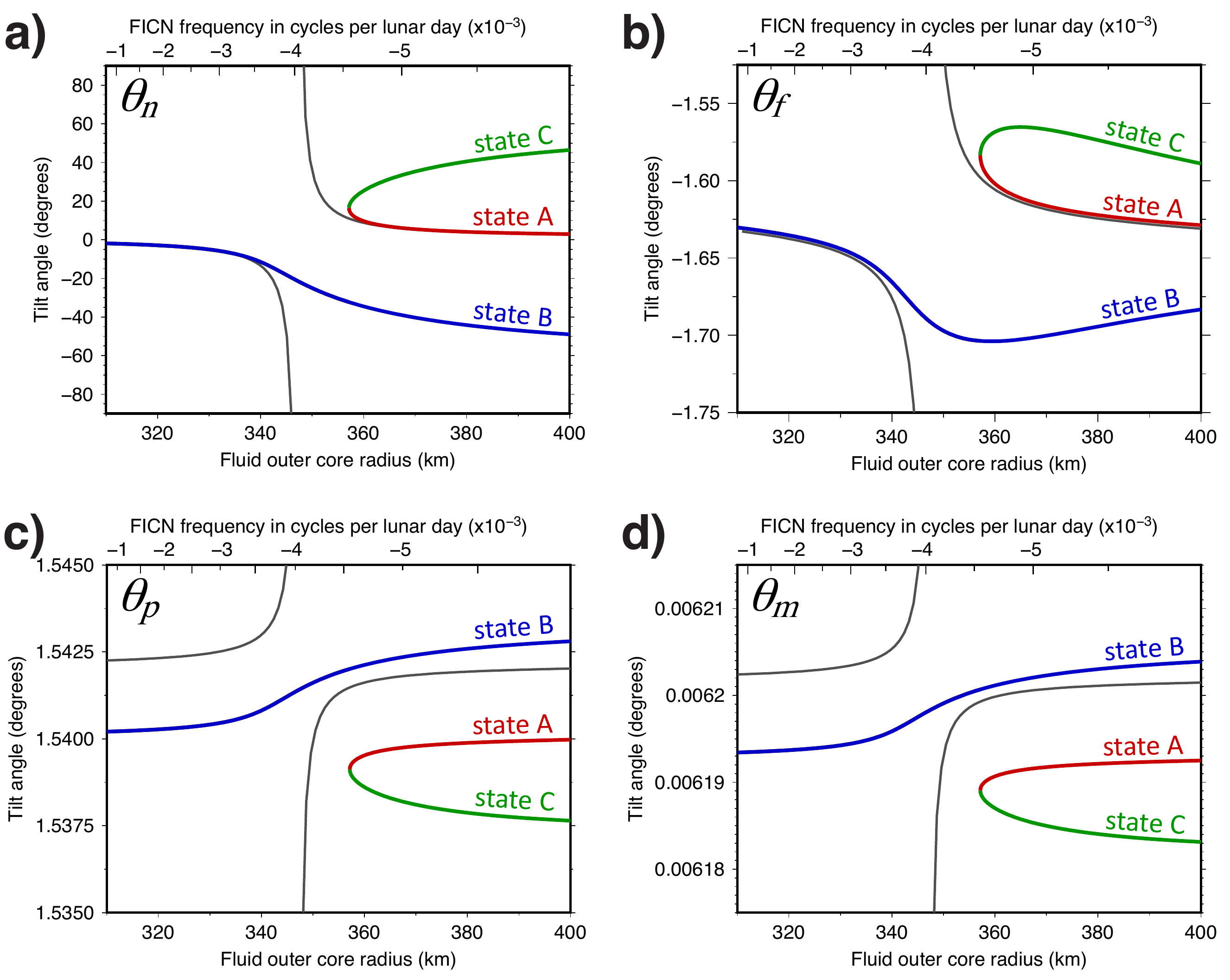}
\caption{\label{fig:res1} The tilt angles (a) $\theta_n$, (b) $\theta_f$, (c) $\theta_p$ and (d) $\theta_m$ as a function of the fluid core outer radius (bottom axis) and FICN frequency (top axis).  The red, blue and green lines correspond to the states A, B and C, respectively, of our general model.  The grey line is the solution of the linear system  in the small-angle limit. $\theta_p$ is measured with respect to the ecliptic normal; $\theta_n$, and $\theta_m$ are measured with respect to the mantle frame; $\theta_f$ is measured with respect to the mantle frame plus $\theta_m$. }
\end{center}
\end{figure}

Fig. \ref{fig:res1} shows $\theta_n$, $\theta_f$, $\theta_p$ and $\theta_m$ obtained from our generalized model and in the small-angle limit.   $\theta_s$ is not shown, as it is virtually identical to $\theta_n$ (the relative difference between the two is of the order of $\delta \omega$).    Results are shown for a Moon model with a crust of thickness $h_c=38.5$ km and density $\rho_c = 2550$ kg m$^{-3}$ \cite[][]{wieczorek13}, an inner core of radius $r_s = 200$ km and density $\rho_s = 7700$ kg m$^{-3}$ \cite[e.g.][]{matsuyama16}, and a range of possible outer core radius between $r_f = 310$ km and 400 km compatible with seismic studies \citep{weber11,garcia11}.  As explained in the previous section, the densities of the fluid core and mantle change for each value of $r_f$, so as to match $M$ and $I_{sm}$; from $r_f= 310$ to $400$ km, $\rho_f$ changes from $7355.7$ to $4772.5$ kg m$^{-3}$, and $\rho_m$ changes from $3376.1$ to $3377.9$ kg m$^{-3}$.

The range of $r_f$ values covered in Fig. \ref{fig:res1} samples different interior Lunar density distributions which in turn samples different frequencies of the FICN, $\omega_{ficn}$.  As predicted by Eq.~\ref{eq:ficncond}, the tilt angle of the inner core should be primarily controlled by $\omega_{ficn}$. The way in which $\omega_{ficn}$ (computed from Eq. \ref{eq:omficn}) changes for each choice of $r_f$ is shown on the top axis in each panels of Fig. \ref{fig:res1}.   For small $r_f$, $\omega_{ficn}$ is slower (in the retrograde direction) than $\delta \omega = \Omega_p/\Omega_o = 4.022 \times 10^{-3}$, the Poincar\'e number, or the retrograde frequency of the forced precession expressed in cycles per Lunar day.  For large $r_f$, the retrograde $\omega_{ficn}$ is instead faster than $\delta \omega$.  

The dominant contribution to $\omega_{ficn}$, as given by Eq.~(\ref{eq:omficn}), is from the gravitational coupling term, $-e_s \alpha_3 \alpha_g$. The change in $\omega_{ficn}$ with $r_f$ shown in Fig. \ref{fig:res1} is a consequence of the change in $\rho_f$ with $r_f$ in our interior models, which results in a change in both $\alpha_3$ (see Eq.~\ref{eq:alpha13}) and $\alpha_g$ (see Eq.~\ref{eq:alphag}).  From $r_f= 310$ to $400$ km, $\alpha_3$ changes from $0.0447$ to $0.3802$, and $\alpha_g$ changes from $105.75$ to $85.76$.  The dynamical ellipticity of the inner core $e_s$ also changes with $r_f$, but the change is modest, from $1.7865 \times 10^{-4}$ to $1.8286 \times 10^{-4}$.

When $\omega_{ficn} = -\delta \omega$, which occurs at $r_f \approx 347$ km in Fig. \ref{fig:res1}, the FICN mode is in perfect resonance with the forcing period.  At that location, solutions in the small-angle limit diverge towards $\pm \infty$. In contrast, solutions from our general model remain finite, even in the proximity of the FICN resonance.  Furthermore, although only one solution is possible when $|\omega_{ficn}| < \delta \omega$, three possible solution branches exist for $|\omega_{ficn}| > \delta \omega$.  This is analogous to the different possible Cassini states of a single-body Moon first highlighted by \cite{peale69} who identified four possible states, numbered 1 to 4. \cite{ward75} showed (his Fig. 2) an example of how states 1, 2 and 4 may have evolved as a function of Earth-Moon distance.  State 3 features a tilt angle larger than $\pm90^\circ$, so a rotation direction opposite to the orbital rotation, and is believed to be unstable when tidal dissipation is taken into account \cite[e.g.][]{peale74}. Currently, the Moon -- or more formally, the outer solid shell made up of its mantle and crust -- occupies state 2, the only state possible when the frequency of the free retrograde precession of the Moon ($\omega_{fp}$) is smaller (in magnitude) than the Poincar\'e number $\delta \omega$.  But in the past when $|\omega_{fp}| > \delta \omega$, states 1, 2, and 4 were all possible solutions.  The number of possible Cassini states, and the angle of mantle precession $\theta_p$ for each, depends essentially on how $\omega_{fp}$ compares with $\delta \omega$ \cite[e.g.][]{peale74}.  

By analogy, the different branches shown on Fig.~\ref{fig:res1} show the different possible Cassini states that are associated with the inner core.   The controlling factor to determine which states are possible, and the angle $\theta_n$ in each of these states, is how the FICN frequency compares with $\delta \omega$ (see Eq.~\ref{eq:ficncond}).  We have labelled these states A, B and C to avoid a possible confusion with the Cassini states associated with the mantle and crust.  State B, the only state possible when $|\omega_{ficn}| < \delta \omega$, features negative values of $\theta_n$: as seen in the Cassini frame, the inner core is tilted away from the mantle, in the direction of the orbit normal.  States A and C, which are only possible when $|\omega_{ficn}| > \delta \omega$, instead have $\theta_n>0$: the inner core is tilted further away than the mantle from the orbit normal.  A state which features $90^\circ < |\theta_n| < 180^\circ$ is also a solution (the analogy of state 3 of the solid shell of the Moon), though we deem such a state impossible as it would feature an inner core rotating in reverse direction with the rest of the Moon.

The point of merging between states A and C (at $r_f \approx 357$ km in Fig. \ref{fig:res1}) correspond to a saddle-point bifurcation.  States A and C exist for $r_f < 357$ km but as purely imaginary solutions, complex conjugates of one another.   Although for $r_f > 357$ km all three states are valid mathematical solutions, state A is preferred because tidal dissipation is expected to drive the system towards its lowest energy state \cite[e.g][]{peale74}.   The inner core core would then be in state B for $r_f < 357$ km and state A for $r_f > 357$ km.  The transition at $r_f \approx 357$ km marks the location of the maximum possible precession angle of the inner core in each of these states.  The solutions shown in Fig. \ref{fig:res1}a suggest that $\theta_n$ could be as large as $17^\circ$ if in state A, or as large as $-33^\circ$ (in the reverse direction) if in state B.   The exact value depends on the FICN frequency of the Moon.

The Cassini state of the inner core manifests itself on the other precession angles.  $\theta_f$ (Fig. \ref{fig:res1}b) shows variations correlated with the variations in $\theta_n$, though much smaller in amplitude.  At the transition between states A and B, $\theta_f$ varies from $-1.59^\circ$ to $-1.71^\circ$, a change in amplitude of $\Delta \theta_f=0.12^\circ$ that is attributable to the inner core.  Likewise, $\theta_p$ and $\theta_m$ (Fig. \ref{fig:res1}c,d) are also adjusted.  At the transition between states A and B, the change in amplitude of $\theta_p$ attributable to the inner core from state A to B is $\Delta \theta_p=0.003^\circ$. Note that in all solutions on Fig. \ref{fig:res1}, $\theta_f$ is always larger in amplitude than $\theta_p$.  In other words, as seen in the Cassini frame, the spin axis of the fluid core is not exactly aligned with the ecliptic normal, but is tilted towards the orbit normal by a small angle of the order of $0.05^\circ$ to $0.17^\circ$ with respect to the ecliptic normal.

Away from the FICN resonance, states A and B converge to the solution in the small angles limit.  In fact, provided $|\theta_n| \le 10^{\circ}$, or equivalently, provided $\omega_{ficn}$ differs from $\delta \omega$ by more than approximately 15\%, the small angle approximation is reasonably accurate. Note that there is a small offset between the general solutions and the small angle approximation solutions of $\theta_p$ and $\theta_m$.  This is caused by the approximations of Eqs. (\ref{eq:app1}-\ref{eq:app3}). Also note that away from the FICN resonance, the solution that we obtain for $\theta_p \approx 1.540^{\circ}$ does not match the observed mantle tilt angle of $1.543^{\circ}$.  This small difference is caused primarily by the omission in our model of the Solar torque acting on the Moon.

\begin{figure}
\begin{center}
\includegraphics[height=8cm]{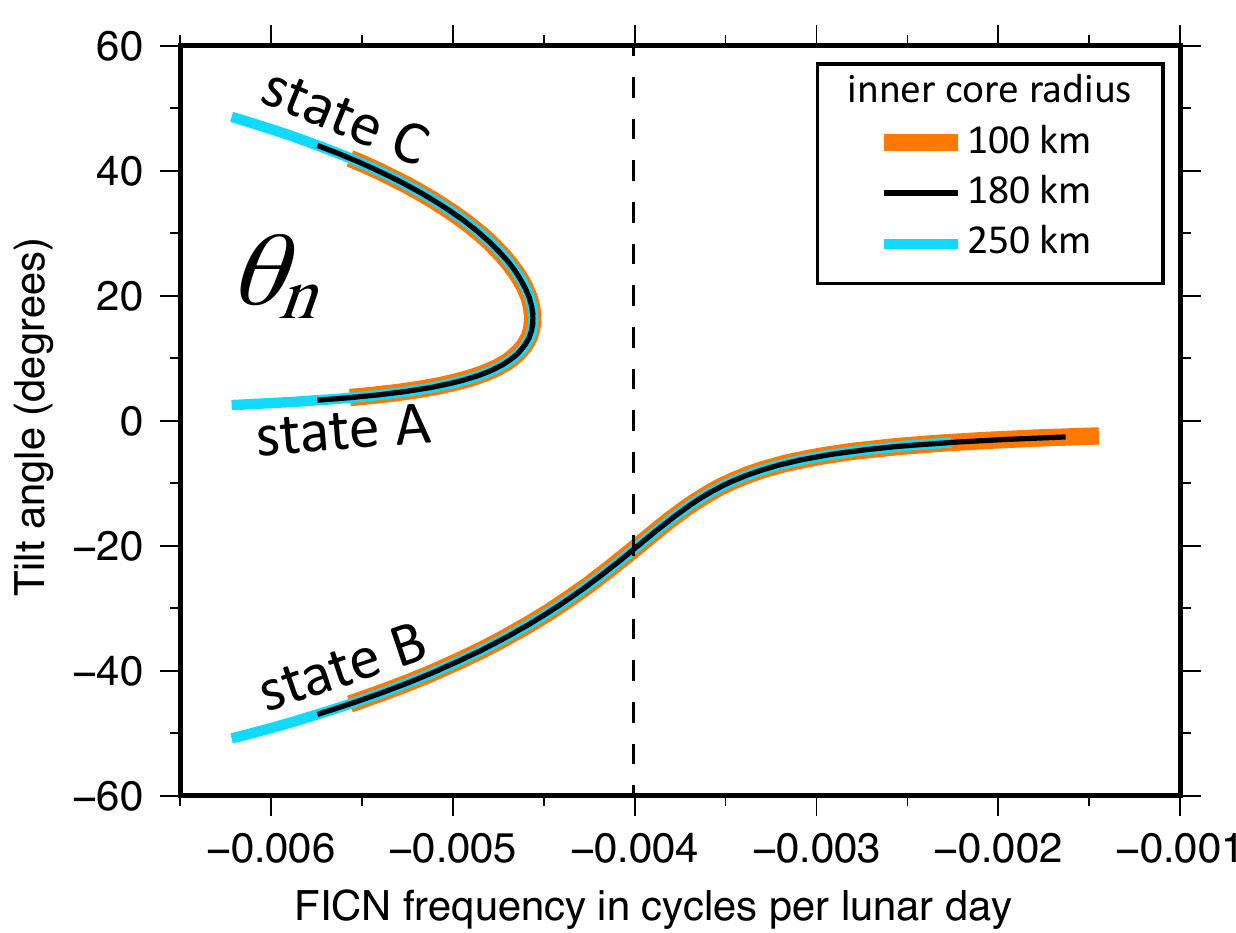}  
\caption{\label{fig:ficn} The tilt angle of the inner core $\theta_n$ as a function of the FICN frequency $\omega_{ficn}$, computed for a range of outer core radii $r_f = [320, 400]$ km and three different choices of inner core radius $r_s$: 100 km (orange), 180 km (black) and 250 km (light blue).  The thickness of each line is varied to reveal that solutions overlie one another.  The location of the FICN resonance is  indicated by the vertical dashed line.}
\end{center}
\end{figure}

According to Eq.~\ref{eq:ficncond}, the tilt angle of the inner core that characterizes its Cassini state depends on the interior density structure of the Moon but only insofar as it influences the frequency of the FICN.  To demonstrate this, Fig. \ref{fig:ficn} shows how $\theta_n$ varies as a function of $\omega_{ficn}$ for three different choices of inner core radii: 100, 180 and 250 km.  In each case, the same range of $r_f = [320, 400]$ km is used.  Although the range of $\omega_{ficn}$ values that is accessed by each choice of $r_s$ is different, the solution for $\theta_n$ versus $\omega_{ficn}$ remains unchanged.  Eq.~\ref{eq:ficncond} provides a very good fit to the variations of $\theta_n$ as a function of $\omega_{ficn}$ shown in Fig.~\ref{fig:ficn}.   

Eq. (\ref{eq:ficncond}) also reveals why the transition from one to three Cassini states is connected to $\omega_{ficn}$.  When the magnitude of $\omega_{ficn}$ is smaller than  $\delta \omega$ (on the right-hand side of the dashed line in Fig.~\ref{fig:ficn}), $\sin(\theta_p+\theta_n)$ must be smaller than $\sin(\theta_n) \cos(\theta_n)$, which is only possible if $\theta_n$ is negative (state B).  Conversely, when the magnitude of $\omega_{ficn}$ is larger than  $\delta \omega$, $\sin(\theta_p+\theta_n)$ must be larger than $\sin(\theta_n) \cos(\theta_n)$.  For $\theta_n$ of the same order as $\theta_p$, this is only possible if $\theta_n$ and $\theta_p$ add up to a larger angle, in other words, if $\theta_n$ is positive (state A on Fig.~\ref{fig:ficn}).   For $\theta_n \gg \theta_p$, Eq. (\ref{eq:ficncond}) becomes 

\begin{equation}
\delta \omega + \omega_{ficn} \cos(\theta_n) = 0\, , \label{eq:ficncondlargen}
\end{equation} 
and this balance is only possible for  $|\omega_{ficn}| > \delta \omega$, and admits a pair of solutions $\pm \theta_n$;  these are the solutions of states B and C on the left-hand side of the dashed line on Fig.~\ref{fig:ficn}.  Eq.~(\ref{eq:ficncond}) also explains why the transition from one to three Cassini states does not occur precisely at the location of the FICN resonance as it involves trigonometric functions of $\theta_p$ and $\theta_n$.  The transition is instead displaced to a larger retrograde value of $\omega_{ficn}\approx-0.00455$ in cycles per Lunar day, or $\omega_{ficn}\approx-2\pi/16.4$ yr$^{-1}$.  

\begin{figure}
\begin{center}
\includegraphics[height=8cm]{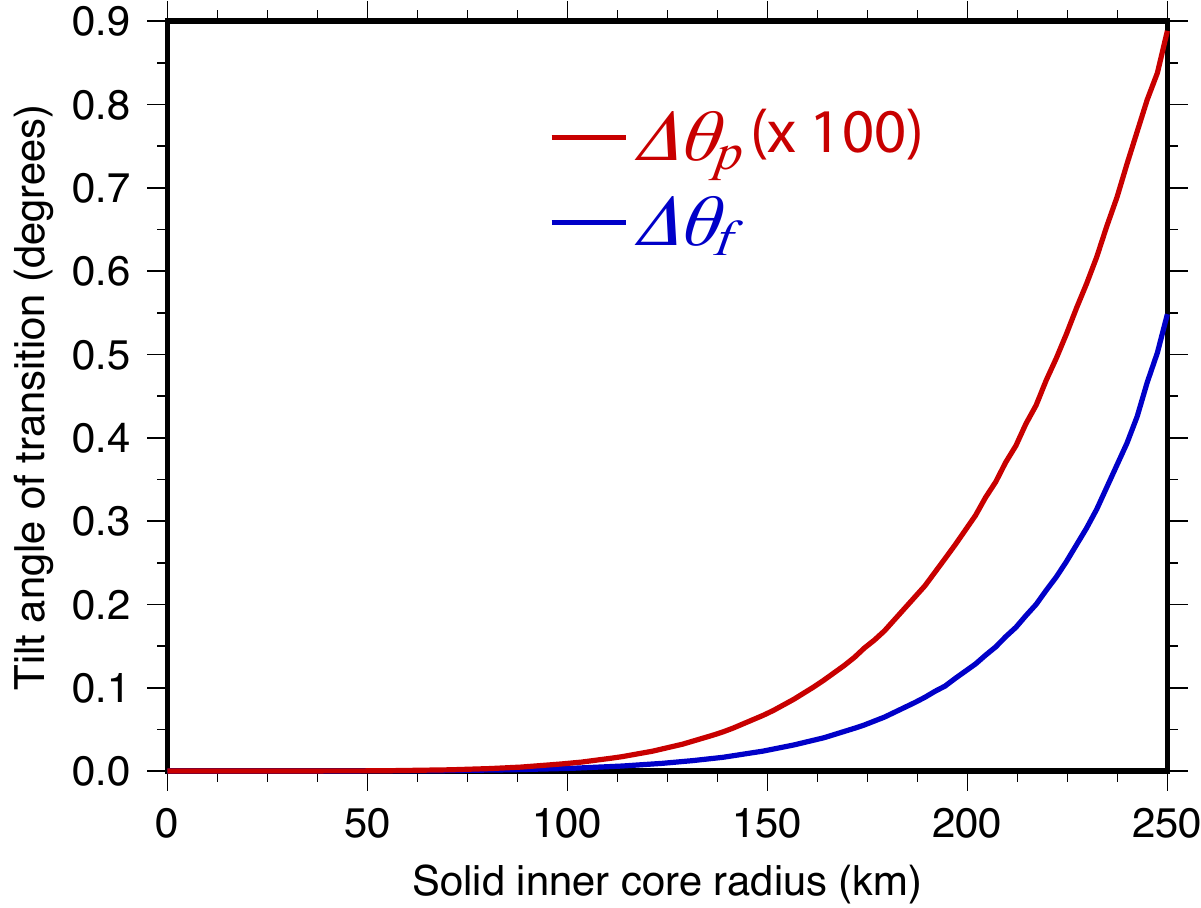}  
\caption{\label{fig:amppf} The amplitude of change of $\theta_p$ (red, multiplied by a factor 100) and $\theta_f$ (blue) at the transition between Cassini states A and B associated with the inner core as a function of inner core radius.  }
\end{center}
\end{figure}

Though the branches of solutions of $\theta_n$ versus $\omega_{ficn}$ are independent of the interior density structure,  it is not the case for $\theta_p$, $\theta_f$ and $\theta_m$.  For the latter three angles, the degree of separation of the solutions into three distinct branches reflects how the mantle and fluid core adjust in response to a tilted inner core. The amplitude of this response depends on the importance of the inner core in the angular momentum balance of the Moon.  To illustrate this, Fig. \ref{fig:amppf} shows how the amplitude of the transition between states A and B for $\theta_p$ and $\theta_f$ (denoted $\Delta \theta_p$ and $\Delta \theta_f$, respectively) changes as a function of inner core radius.  The larger the inner core, the more important its influence is in the angular momentum dynamics of the Moon.  Therefore, the greater the manifestation of the Cassini state associated with the inner core is on $\theta_f$ and $\theta_p$.  For an inner core smaller than 100 km, the Cassini state of the inner core has a vanishingly small influence on $\theta_f$ and $\theta_p$.  But for an inner core as large as 250 km, $\Delta \theta_p$ gets close $0.01^\circ$.  This implies that, for a large inner core, the observed mantle tilt angle of $1.543^\circ$ could include a small though non-negligible contribution from the inner core, the exact amount depending on the inner core size and how close to resonance the frequency of the FICN is. A large inner core has a more dramatic influence on $\theta_f$ because the moment of inertia of the fluid core is much smaller than that of the solid shell.  For an inner core radius of 250 km, $\Delta \theta_f$ gets as large as approximately $0.5^\circ$. 

\begin{figure}
\begin{center}
\includegraphics[height=8cm]{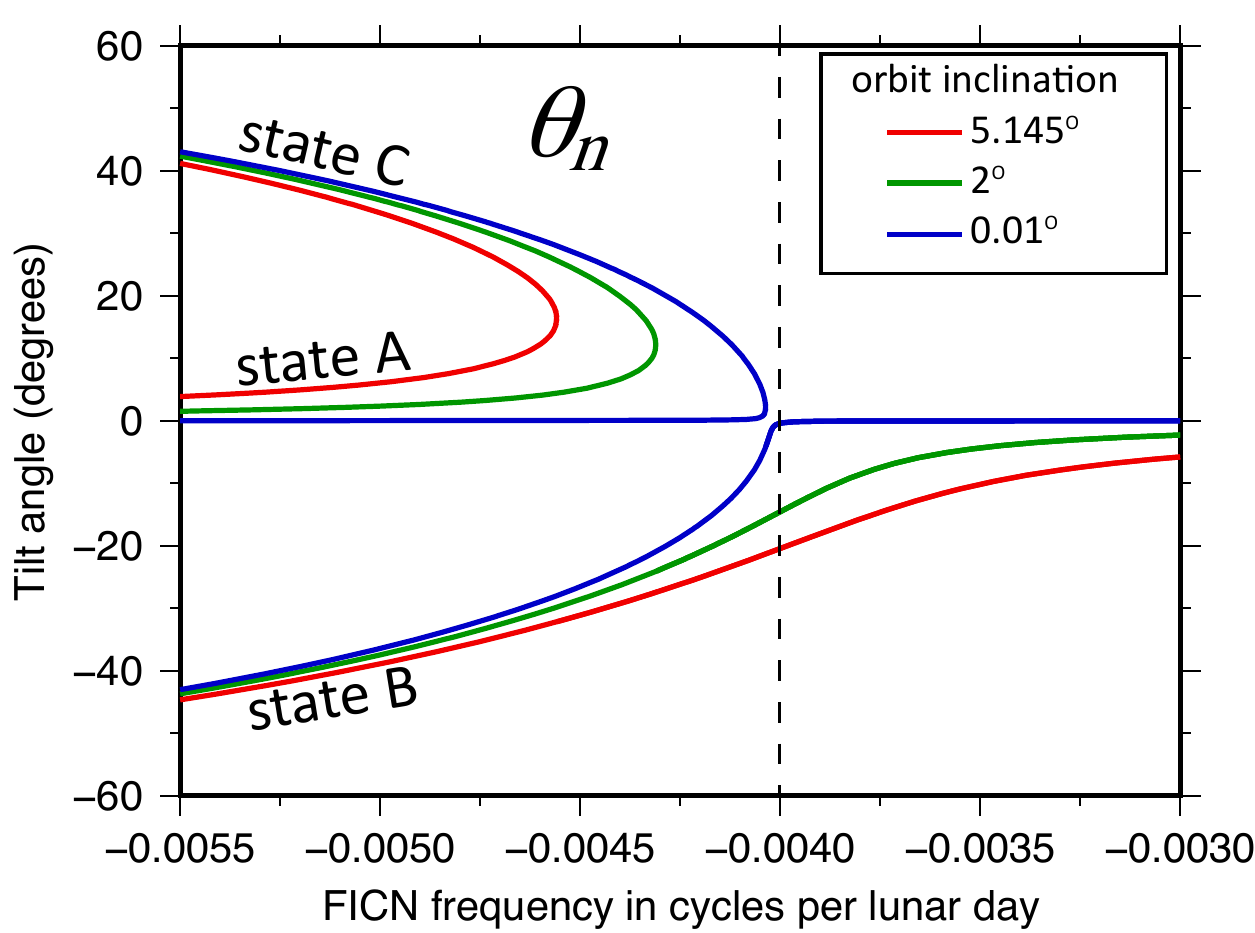}  
\caption{\label{fig:incl} The tilt angle of the inner core $\theta_n$ as a function of the FICN frequency $\omega_{ficn}$ for three different choices of the orbital inclination: $I=5.145^\circ$ (red);  $I=2^\circ$ (green);  $I=0.01^\circ$ (blue). The location of the FICN resonance is indicated by the vertical dashed line.}
\end{center}
\end{figure}

Lastly, it is instructive to show how the shape of the branches of solution change when the geometry of torque by Earth is modified.  Fig. \ref{fig:incl} shows how the branches of solutions of $\theta_n$ are altered for three different choices of the orbital inclination: $I=5.145^\circ$, $I=2^\circ$ and $I=0.01^\circ$.  As $I$ approaches zero, the point of merging between states A and C approaches the location of the FICN resonance and the transition from three to one state approaches the shape of a pitchfork bifurcation. Eq. (\ref{eq:ficncond}) remains a very good approximation to the solutions shown in Fig. \ref{fig:incl}; the change in the shape of the solutions occurs because as $I \rightarrow 0$, $\theta_p \rightarrow 0$. 

\section{Discussion and conclusion}

We showed in this study that the angle of tilt of the inner core of the Moon that characterizes its Cassini state depends on the frequency of the FICN, $\omega_{ficn}$. More specifically, that it depends on how the magnitude of $\omega_{ficn}$ compares with the Poincar\'e number $\delta \omega = \Omega_p/\Omega_o$.   For the present-day Moon, with its rotation rate of $\Omega_o = 2\pi/27.322$ days$^{-1}$ and precession frequency $\Omega_p=2\pi/18.6$ yr$^{-1}$, we can cast our results in terms of a comparison between $\omega_{ficn}$ and $\Omega_p$ both given in frequency units.  Denoting the tilt angle of the inner core with respect to the mantle as $\theta_n$, our results show that if $|\omega_{ficn}| \gg \Omega_p$, $\theta_n$ is positive but approaches zero (state A on Fig.~\ref{fig:ficn} and see also DW16): the inner core remains closely aligned with mantle.   If instead  $|\omega_{ficn}| \ll \Omega_p$, $\theta_n$ is negative and small (state B on Fig.~\ref{fig:ficn}), but does not converge to zero: a small misalignment with the mantle remains (see DW16).  In between these two extremes, $\theta_n$ can be large, as the inner core precession is resonantly amplified by the proximity of $\omega_{ficn}$ to the forcing frequency $\Omega_p$.  Assuming the lowest energy state is favoured, the largest positive $\theta_n$ in state A is $17^{\circ}$ and the largest negative $\theta_n$ in state B is $-33^{\circ}$. The transition between these two extremes does not occur exactly at $\omega_{ficn} = -\Omega_p$, but instead at $\omega_{ficn} = -2\pi/16.4$ yr$^{-1}$.

\begin{figure}
\begin{center}
\includegraphics[height=8cm]{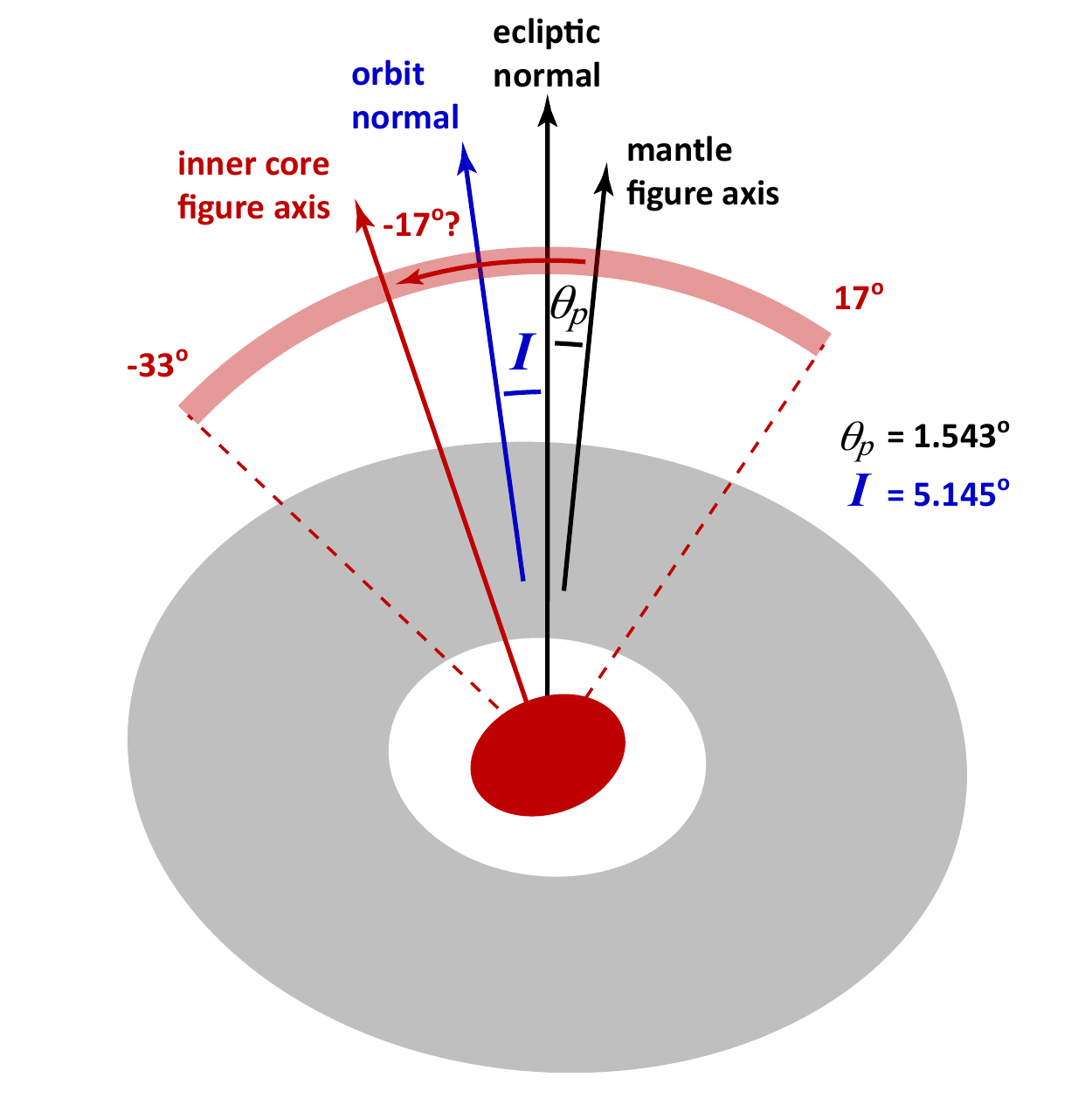}  
\caption{\label{fig:siccass} The Cassini state of the inner core, as seen in the Cassini frame.  The red shaded arc shows the possible range of inner core precession angles, from $+17^\circ$ to $-33^\circ$, measured with respect to the mantle figure axis.  Adopting a Lunar interior density model close to that of \cite{matsuyama16} gives a precession angle of $-17^\circ$.  Angles, ellipticities and region thicknesses are not drawn to scale.}
\end{center}
\end{figure}

The precise angle of tilt of the inner core depends then on the knowledge of $\omega_{ficn}$, which in turn depends on the knowledge of the interior structure of the Moon.   The uncertainty in the latter is large enough that a considerable range of $\omega_{ficn}$ values are possible, from approximately half to twice as large as $\Omega_p$ (DW16).  This places $\omega_{ficn}$ within the resonance band of the forced 18.6 yr precession.  Consequently, we expect the inner core to be substantially misaligned with the mantle.  As an illustrative example, let us calculate the FICN frequency for one possible interior structure model.  We pick as a basis one of the model presented in \cite{matsuyama16}, specifically the model in their Table 2 without a low velocity layer at the bottom of the mantle, and constrained to match the Lunar mass, the moment of inertia, and the observed values of $k_2$ and $h_2$ (and the model for which $h_2$ is derived from LLR).  Using the central values for the density and radius of each layer, the central values for the densities of the inner core and crust, we find $\rho_m$= 3358 kg/m$^3$ and $\rho_f$= 5878.6 kg/m$^3$ from fitting $I_{sm}$ and $M$ by the method described in section 3.1.  These are compatible with the range of values of $\rho_m$ and $\rho_f$ given in Table 2 of \cite{matsuyama16}.  The FICN frequency that we calculate for this specific model is $\omega_{ficn} = - 2 \pi/19.48$ yr$^{-1}$.  According to our adopted denomination, the inner core would be in state B and its tilt angle (predicted by Eq.~\ref{eq:ficncond}) would be approximately $-17.16^\circ$.  That is, as seen in the Cassini frame, the inner core is offset from the mantle axis by $\sim 17^\circ$, towards the ecliptic normal (Fig.~\ref{fig:siccass}). 

However, because of the proximity of $\omega_{ficn}$ to $\Omega_p$, the precise value of the inner core tilt is very sensitive to small changes in the interior density structure. So we must emphasize that the uncertainty on the inner core precession angle remains large, and it could take any values between $-33^\circ$ and $17^\circ$ (Fig.~\ref{fig:siccass}).  Additionally, we have assumed here the mantle to be a simple one layer model.  The latest  models of the Lunar interior allow for a higher density, low seismic velocity layer at the bottom of the mantle \cite[][]{weber11,matsumoto15,matsuyama16}, which would further influence the exact frequency of the FICN, and thus the tilt angle of the inner core.  

Our results also indicate that if the inner core is sufficiently large, it can contribute to the observed tilt angle of the solid outer shell of the Moon of $\theta_p=1.543^\circ$.   By exactly how much depends on the inner core size and how close the FICN frequency is to resonance.  Because the inner core tilt can be either positive (if in state A) or negative (if in state B), it can lead to either a negative or a positive contribution to $\theta_p$, respectively.  For an inner core as large as 250 km, this contribution could be of the order of $\pm0.005^\circ$. Conversely, this implies that parameters inferred from fitting the observed $\theta_p$ can take different numerical values when determined on the basis of a Moon model with a large inner core versus one with a small or no inner core.  This is the case notably for the parameter $\beta$ given in Eq. (\ref{eq:betagamma}).  Since $\beta$ involves the moments of inertia, a change in its numerical value corresponds to a different constraint on the Moon's interior structure.  Consequently, interior models constructed on the basis of this constraint would then also be altered. 

Likewise, a large inner core can induce a substantial change in the orientation of the rotation vector of the fluid core.  The latter is typically assumed to be closely aligned with the ecliptic because the frequency of the FCN is much smaller than the forcing precession frequency \cite[][]{poincare10,goldreich67,meyer11}.  But as we have shown here, if the FICN frequency is very close to the resonance, a large inner core can entrain a significant misalignment of the fluid core spin axis from the ecliptic of the order of $\pm0.2^\circ$.

One improvement to our model would be to include elastic deformations, which we have neglected.  The  prediction of the mantle tilt angle in our model is $\theta_p \approx 1.540^\circ$, off by approximately 0.19\% from the observed tilt of $\theta_p = 1.543^\circ$, dominantly because we have neglected the torque from the Sun. Including elastic deformations would not contribute to a large additional correction to $\theta_p$, but changes in the tilt angles of the inner core and fluid core could be more important.  The largest force acting on a tilted inner core is from gravitational coupling with the mantle and fluid core.  Elastic deformations would act to realign the inner core with the mantle, so would lead to a decrease in the inner core tilt angle. The range of possible inner core tilt angles quoted above, from $-33^\circ$ to $17^\circ$, could be slightly diminished.  

Perhaps more importantly, viscous relaxation within the lower portion of the mantle \cite[][]{harada14} or within the inner core may also substantially alter our results.  In particular, an inner core that can deform viscously would realign its shape to match the surface of hydrostatic equilibrium imposed by the mantle gravity field. As was shown in DW16, if the viscous relaxation timescale of the inner core is of the order of one Lunar day, gravitational coupling with the mantle would prevent a misalignment of the  inner core of more than $1^\circ$. 

\begin{figure}
\begin{center}
\includegraphics[height=6cm]{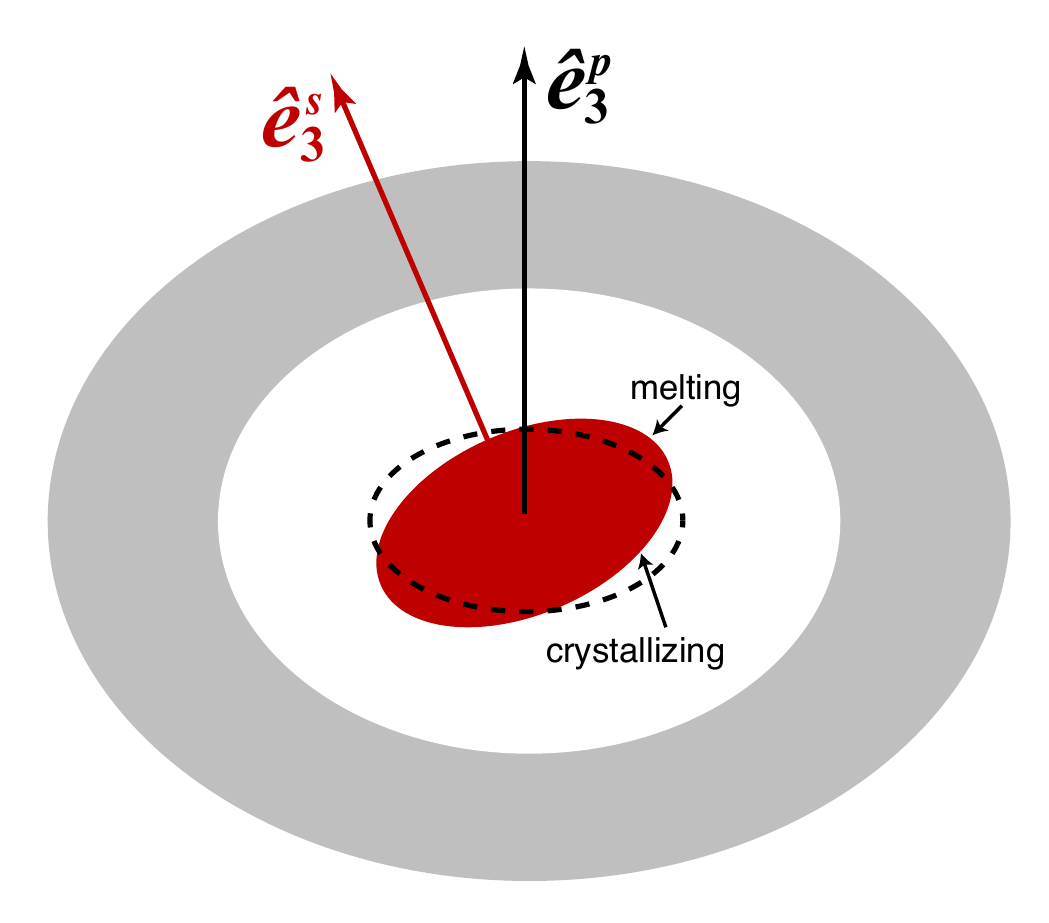}  
\caption{\label{fig:melt} The surface of a tilted ellipsoidal inner core (red) with respect to the mantle figure (grey) is misaligned with the liquidus temperature (dashed line).  Regions of the inner core boundary at a larger (smaller) radius than the liquidus experience melting (crystallizing).  Ellipticities and region thicknesses are not drawn to scale.}
\end{center}
\end{figure}

Another process that acts to realign the ICB to the surface of hydrostatic equilibrium imposed by the mantle is melting and crystallizing at the top of the inner core.  Within the fluid core, hydrostatic equilibrium implies that surfaces of constant gravitational potential, density and pressure are all aligned.  Linked to pressure and density by an equation of state, surfaces of constant temperature follow the same alignment.  The ICB marks the transition from the solid to liquid phase of the core Fe-alloy, so at equilibrium its temperature should coincide with the liquidus (melting temperature).  A tilted ellipsoidal inner core however has its ICB misaligned from the liquidus (Fig.~\ref{fig:melt}).  Parts of the ICB that are at a higher radius than the liquidus undergo melting, parts that are at a lower radius are the seat of crystal growth.  A tilted inner core is precessing at frequency $\omega \Omega_o$ in the frame of the mantle, so over the course of one Lunar orbit around Earth, a given point on the ICB goes through a cycle of melting and crystallizing.  At each moment in this cycle, melt or growth of the ICB is always directed towards an alignment with the liquidus.  Over a long period of time, this should act to realign the shape of the ICB with the liquidus, and thus to realign the figure axis of the inner core with that of the mantle.

These considerations have not been taken into account in our model.  A more proper determination of the tilt angle of the inner core would involve a balance between two characteristic timescales: the timescale of realignment of the ICB to the liquidus by melt and growth versus the timescale for the inner core to assume the tilt angle of its Cassini state when starting from an alignment with the mantle.

The important point to stress is that both viscous relaxation and the process of melting and solidification of the ICB act to reduce the amplitude of the inner core tilt predicted by our simple model.  Indeed, this may be part of the reason why the periodic degree 2 and order 1 gravity signal associated with the inner core, which is expected to be above detection level \cite[][]{williams07,zuber13}, has so far remained undetected \cite[][]{williams15b}.  In fact, for the interior model with a predicted inner core tilt of $-17^\circ$ presented above, this signal should be of the order of $2-3 \times 10^{-10}$, large enough enough to be detected.  The non-detection of this gravity signal may then reflect the importance of viscous relaxation or melting/solidification acting to reduce the inner core tilt.  Alternately, the non-detection may be because the inner core is too small, the density contrast at the ICB is too small, or that the FICN frequency is not near the resonance so the inner core tilt is too small.

We have applied our model of the Cassini state of the inner core to the present-day orbital configuration of the Moon.  But the orbit of the Moon has evolved with time. The rotation rate of the 1:1 spin-orbit resonance has decreased as the Earth-Moon distance increased.  This implies a change in the FICN frequency with time.  Likewise, the orbital precession frequency, and thus the forced precession frequency, is also changing in time. For a fixed Lunar interior structure, the changing ratio of $\omega_{ficn}$ to $\Omega_p$ implies that the tilt angle of the inner core is expected to also change in time.  It is even possible that a transition from one Cassini state to another may have occurred in the past, or will occur in the future.  In fact, Fig. 2 of \cite{ward75} illustrates precisely this, showing how the different Cassini states associated with the solid outer shell of the Moon have evolved as a function of the Earth-Moon distance.  The transition between states 1 ($\theta_p<0$) and 2 ($\theta_p>0$) marks the resonance crossing of the free precession of the Lunar mantle in space.  

These considerations are important in regards to the origin of the past Lunar dynamo \cite[e.g.][]{weiss14}.  One suggestion that has been proposed is that the dynamo may have been sustained by mechanical forcing from differential rotation at the CMB \cite[][]{williams01,dwyer11}, when the tilt angle of the spin symmetry axis of the mantle with respect to the ecliptic was larger \cite[][]{ward75}.  In this model, the fluid core spin axis is assumed to be perfectly aligned with the ecliptic normal.  But as we showed in our study, the fluid core spin axis may be sufficiently offset from the ecliptic if the inner core is large and the FICN frequency is close to resonance.  Furthermore, in the past, the FCN frequency  was larger, principally because of the faster rotation rate of the Moon \cite[but also if this resulted in a greater CMB ellipticity, e.g.][]{meyer11}, and thus the FCN was closer to being in resonance with the forced precession frequency.  Consequently, even for a small or no inner core, the offset of the fluid core spin axis with the ecliptic was larger in the past, thus enhancing the power available to drive a mechanical Lunar dynamo.  The factor of enhancement depends on the Lunar interior model and on the details of the evolution of the Lunar orbit.  We plan to investigate this in a follow up study.

Since, as we have illustrated in our study, the spin vectors of the solid and fluid cores are likely misaligned, it follows then that the resulting differential rotation at the ICB may also potentially lead to a dynamo by mechanical stirring.  Clearly, if this mechanism is possible, then the differential velocity at the ICB at present is too small for dynamo action, either because the inner core is too small, or because the differential precession angle between the inner core and the spin axis of the fluid core is too small, or both.  But if it is because of the latter, the different ratio between $\omega_{ficn}$ and $\Omega_p$ in the past may have lead to a much larger inner core tilt -- even possibly a resonance crossing of the FICN -- and a sufficiently large differential rotation at the ICB for dynamo action.  Whether this may have occurred depends on the evolution of the Lunar orbit parameters, the Lunar interior structure and on how the latter may have evolved (for instance by inner core growth).  We also plan to investigate this in a follow up study.

The melting and solidification cycle of a tilted inner core discussed above also introduce another possibility for dynamo action.  This corresponds to a degree 2 order 1 spherical harmonic source/sink pattern of latent heat at the ICB, rotating in the retrograde direction with a period of one Lunar day.  The larger the tilt angle of the inner core, the greater the amplitude of the latent heat source/sink.  Is it possible that a larger inner core tilt angle in the past lead to dynamo action through convection in the fluid core powered by this process?

Finally, our model can be easily adapted to other moons and planets in a Cassini state, such as Mercury \cite[][]{peale16}, or to other bodies that have a solid-liquid-solid structure, such as the icy satellites of Jupiter and Saturn \cite[e.g.][]{baland12}.  Elastic deformations are important in the thin mantle of Mercury \cite[e.g.][]{mazarico14} and for flexible icy shells \cite[e.g.][]{vanhoolst13} and in order to capture correctly the precession dynamics of these bodies, they will have to be incorporated. 



\appendix 
\section{Description of the Lunar orbit, Lunar rotation and references frames used in our model}

The Moon is rotating around Earth on an eccentric orbit inclined by an angle $I=5.145^\circ$ with respect to the ecliptic plane.  This orbital plane is precessing about the ecliptic normal in a retrograde direction with a frequency of $\Omega_p = 2\pi/18.6$ yr$^{-1}$.  To describe the position of the Moon as it orbits about the Earth, we define a coordinate system attached to the inertial reference frame, centred on Earth, and specified by unit vectors $(\mitbf{\hat{e}_1}, \mitbf{\hat{e}_2}, \mitbf{\hat{e}_3})$.  Direction $\mitbf{\hat{e}_3}$ is aligned with the normal to the ecliptic.  The normal to the orbital plane, defined by a normal unit vector $\mitbf{\hat{e}_3^I}$, is then precessing about $\mitbf{\hat{e}_3}$ at frequency $-\Omega_p$.

\begin{figure}
\begin{center}
\includegraphics[height=5cm]{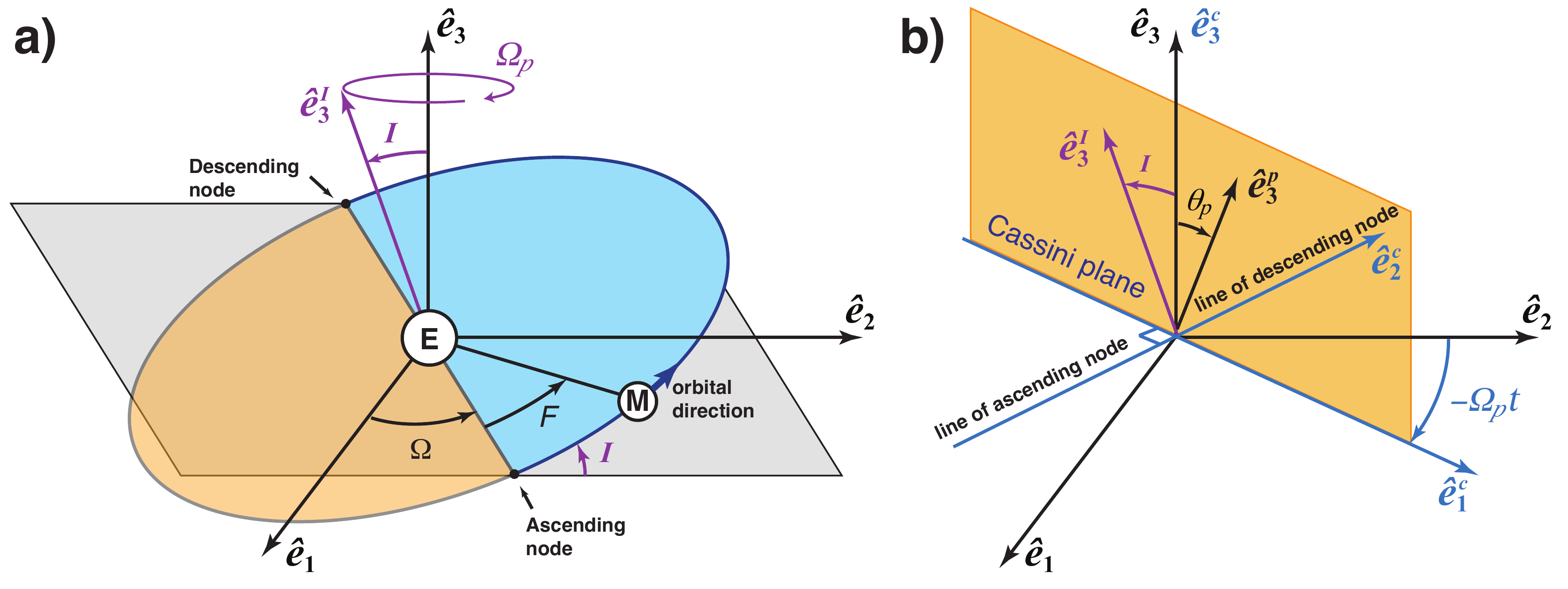}  
\caption{\label{fig:app1} a) The orbit of the Moon (M) around Earth (E) as seen in the inertial frame $(\mitbf{\hat{e}_1}, \mitbf{\hat{e}_2}, \mitbf{\hat{e}_3})$. The normal to the orbital plane is defined by $\mitbf{\hat{e}_3^I}$ and is offset from $\mitbf{\hat{e}_3}$ by an angle $I=5.145^\circ$. $\mitbf{\hat{e}_3^I}$ precesses about $\mitbf{\hat{e}_3}$ in a retrograde direction at frequency $\Omega_p = 2\pi/18.6$ yr$^{-1}$.  $F$ is the mean angle from the orbit's ascending node.  $\Omega$ is the longitude of the ascending node with respect to $\mitbf{\hat{e}_1}$.  The blue (orange) shaded region indicates portions of the orbit when the Moon is above (below) the ecliptic plane, the latter being represented by the grey shade. b) The Cassini frame $(\mitbf{\hat{e}^c_1}, \mitbf{\hat{e}^c_2}, \mitbf{\hat{e}^c_3})$ is rotating at frequency $-\Omega_p$ about $\mitbf{\hat{e}_3}=\mitbf{\hat{e}^c_3}$ with respect to the inertial frame, with $\mitbf{\hat{e}^c_2}$ aligned with the line of the descending node.  The symmetry axis of the mantle $\mitbf{\hat{e}_3^p}$ is offset from $\mitbf{\hat{e}_3}$ by $\theta_p=1.543^\circ$.  Both $\mitbf{\hat{e}_3^I}$ and $\mitbf{\hat{e}_3^p}$ remain in the Cassini plane, the plane defined by $\mitbf{\hat{e}^c_1}$ and $\mitbf{\hat{e}^c_3}$ delimited by the orange shaded region.  a) and b) do not correspond to the same snapshot in time.}
\end{center}
\end{figure}

As shown in Fig.~\ref{fig:app1}a, the position of the Moon is described by an angle $F$, the mean angle from the orbit's ascending node, and by an angle $\Omega$, the longitude of the ascending node with respect to $\mitbf{\hat{e}_1}$.  The rate of change of $\Omega$ is related to the precession frequency by $\frac{d\Omega}{dt} = -\Omega_p$. The time it takes for the Moon to complete one orbit with respect to the inertial frame is defined as the sidereal period and is equal to 27.322 days.  The sidereal frequency is equal to the mean motion, $n= 2\pi/27.322$ day$^{-1}$.  Since the Moon is in a tidally locked 1:1 spin-orbit resonance, the rate of the Moon's rotation around itself averaged over one orbit is closely related to $n$, though not exactly equal, as we develop below.  

Because the orbit is precessing, the Moon does not return to the same point in inertial space after one sidereal period.  The time it takes for the Moon to return to the ascending node of the orbit is slightly shorter than the sidereal period, and is equal to 27.212 days. Defining this orbital frequency by $\Omega_c= 2\pi/27.212$ day$^{-1}$, the mean rate of change of $F$ averaged over one orbit is related to $\Omega_c$ by $\frac{dF}{dt} = \Omega_c$. The mean motion is linked to $\Omega_c$ and $\Omega_p$ by $n=\Omega_c-\Omega_p$.

The half-period modulation of the gravitational torque by Earth over one orbit and the eccentricity of the orbit lead to small latitudinal and longitudinal librations of the Moon in space.  These are neglected in our study, as we focus on the long timescale equilibrium described by the Cassini state.  In other words, in the description of the Cassini state that follows, even when not specifically stated, we always consider quantities that are averaged over one orbit.

The Moon is in a Cassini state, which describes the fact that the symmetry axis (defined by a unit vector $\mitbf{\hat{e}^p_3}$), though inclined by $\theta_p=1.543^{\circ}$ with respect to $\mitbf{\hat{e}_3}$, remains co-planar with both $\mitbf{\hat{e}_3}$ and $\mitbf{\hat{e}^I_3}$.  The plane containing all three vectors is rotating with frequency $-\Omega_p$ about $\mitbf{\hat{e}_3}$ with respect to the inertial frame. This description is only valid when the orientation of $\mitbf{\hat{e}^p_3}$ is averaged over one orbit, which is assumed in our discussion.  To describe the Cassini state, it is convenient to introduce a second reference frame which we refer to as the Cassini frame.  The Cassini frame is specified by unit vectors $(\mitbf{\hat{e}^c_1}, \mitbf{\hat{e}^c_2}, \mitbf{\hat{e}^c_3})$, with $\mitbf{\hat{e}^c_3}$ aligned with the ecliptic normal ($\mitbf{\hat{e}^c_3}=\mitbf{\hat{e}_3}$), and is rotating with frequency $-\Omega_p$ about $\mitbf{\hat{e}_3}$ with respect to the inertial frame (Fig.~\ref{fig:app1}b). The orientation of the Cassini frame is chosen such that direction $\mitbf{\hat{e}^c_2}$ remains aligned with the line of the descending node of the orbit on the ecliptic plane.   It is convenient to place the origin of the Cassini frame at the centre of the Moon.  Setting an alignment $\mitbf{\hat{e}^c_1} = \mitbf{\hat{e}_2}$ at time $t=0$, the  relationship between the Cassini and inertial reference frames is expressed by

\begin{subequations}
\begin{align}
\mitbf{\hat{e}^c_1} & =\phantom{-}\sin (-\Omega_p t) \mitbf{\hat{e}_1} \,  + \, \cos (-\Omega_p t) 
\mitbf{\hat{e}_2} \, , \label{eq:ec1toe} \\
\mitbf{\hat{e}^c_2} & =  - \cos (-\Omega_p t) \mitbf{\hat{e}_1} \, + \, \sin (-\Omega_p t) \mitbf{\hat{e}_2} \, ,\\
\mitbf{\hat{e}^c_3} & =\phantom{-}\mitbf{\hat{e}_3} \, .
\end{align}
\label{eq:ectoe}
\end{subequations}

As viewed in the Cassini frame, both the direction of the orbit normal $\mitbf{\hat{e}_3^I}$ and the symmetry axis $\mitbf{\hat{e}^p_3}$ remain at fixed positions. It is convenient to refer to the plane defined by  $\mitbf{\hat{e}^c_1}$ and $\mitbf{\hat{e}^c_3}$ as the ``Cassini plane'' (Fig. ~\ref{fig:app1}b).  Because the Moon possesses a fluid and (most likely) a solid core, formally $\mitbf{\hat{e}^p_3}$ represents the symmetry axis of the mantle only.  The orientation of the symmetry axis of the inner core, denoted by $\mitbf{\hat{e}^s_3}$ also lies on the Cassini plane, and also remains at a fixed position in the Cassini frame (see Fig.~\ref{fig:rotframes}a of the main text).  

We define the rotation vector of the Moon's mantle by $\mitbf{\Omega}$.  The vector $\mitbf{\Omega}$  also lies on the Cassini plane, though it is not aligned exactly with the symmetry axis $\mitbf{\hat{e}^p_3}$ but is offset by a small angle $\theta_m$ (see Fig.~\ref{fig:rotframes}b of the main text).   To preserve a synchronous rotation, $\mitbf{\Omega}$ as seen in the Cassini frame is given by

\begin{equation} 
\mitbf{\Omega} = \Big[ - \Omega_p + \Omega_c \cos (\theta_p) \Big]  \mitbf{\hat{e}^c_3} + \Omega_c \sin (\theta_p) \,  \mitbf{\hat{e}^c_1} \, , \label{eq:rot_cass}
\end{equation}
and, upon using Eqs. (\ref{eq:ectoe}), by

\begin{equation} 
\mitbf{\Omega} = \Big[ - \Omega_p + \Omega_c \cos (\theta_p) \Big] \mitbf{\hat{e}_3} + \Omega_c \sin (\theta_p)  \Big[ \sin (-\Omega_p t)  \mitbf{\hat{e}_1} + \cos (-\Omega_p t) \mitbf{\hat{e}_2} \Big] \, . \label{eq:rot_inertial}
\end{equation}
when seen in the inertial frame.

The model of the rotational dynamics of the Moon that we develop in the main text is defined with respect to a reference frame attached to the rotating mantle.  We must then express how this reference frame is connected to the inertial and Cassini frames defined above.  Let us define the mantle frame by unit vectors $(\mitbf{\hat{e}^p_1}, \mitbf{\hat{e}^p_2}, \mitbf{\hat{e}^p_3})$.  We have already defined $\mitbf{\hat{e}^p_3}$ to be aligned with the maximum (polar) moment of inertia of the mantle.  $\mitbf{\hat{e}^p_1}$ and $\mitbf{\hat{e}^p_2}$ are aligned, respectively, with the minimum and intermediate moments of inertia (both in equatorial directions).  As seen in the Cassini frame, although $\mitbf{\hat{e}^p_3}$ remains at a fixed orientation, $\mitbf{\hat{e}^p_1}$ and $\mitbf{\hat{e}^p_2}$ are time-dependent because the Moon is rotating about itself.  This is depicted in Figs.~\ref{fig:rotframes}a,b of the main text. 

As seen in the Cassini frame, the time it takes for $\mitbf{\hat{e}^p_1}$ and $\mitbf{\hat{e}^p_2}$ to complete one full rotation must coincide with the time it takes for these vectors to return to the same alignment with respect to Earth.  In other words, the rate of rotation of $\mitbf{\hat{e}^p_1}$ and $\mitbf{\hat{e}^p_2}$ about $\mitbf{\hat{e}^p_3}$ is equal to the orbital frequency $\Omega_c$. 

Setting an alignment $\mitbf{\hat{e}^p_2} = \mitbf{\hat{e}^c_2}$ at time $t=0$, the time-dependent orientation of the mantle axes as seen in the Cassini frame is expressed by

\begin{subequations}
\begin{align}
\mitbf{\hat{e}^p_1} & =\phantom{-} \cos (\theta_p)  \cos (\Omega_c t) \mitbf{\hat{e}^c_1} \, + \, \sin (\Omega_c t) \mitbf{\hat{e}^c_2} \, - \, \sin (\theta_p)  \cos (\Omega_c t) \mitbf{\hat{e}^c_3} \, , \\
\mitbf{\hat{e}^p_2} & = - \cos (\theta_p)  \sin (\Omega_c t) \mitbf{\hat{e}^c_1} \, + \, \cos (\Omega_c t) \mitbf{\hat{e}^c_2} \, + \, \sin (\theta_p)  \sin (\Omega_c t) \mitbf{\hat{e}^c_3}\, , \\
\mitbf{\hat{e}^p_3} & =\phantom{-} \cos (\theta_p)  \mitbf{\hat{e}^c_3} \,  + \, \sin (\theta_p)  \mitbf{\hat{e}^c_1} \, .
\end{align}
\label{eq:eptoec}
\end{subequations}
Using Eqs. (\ref{eq:ectoe}), the time-dependent orientation of the mantle axes as seen in the inertial frame is expressed by

\begin{subequations}
\begin{align}
\mitbf{\hat{e}^p_1}  = & \phantom{-} \, \:  \Big[  \cos (\theta_p)  \cos (\Omega_c t) \sin (-\Omega_p t) \, - \,  \sin (\Omega_c t)  \cos (-\Omega_p t) \Big] \mitbf{\hat{e}_1}  \nonumber\\
&+\Big[ \cos (\theta_p)  \cos (\Omega_c t) \cos (-\Omega_p t) \, + \,  \sin (\Omega_c t)  \sin (-\Omega_p t) \Big] \mitbf{\hat{e}_2}  \nonumber\\
& -\sin (\theta_p)  \cos (\Omega_c t) \mitbf{\hat{e}_3} \, ,\\
\mitbf{\hat{e}^p_2}  =& \phantom{-} \, \: \Big[ - \cos (\theta_p)  \sin (\Omega_c t) \sin (-\Omega_p t) \, - \,  \cos (\Omega_c t)  \cos (-\Omega_p t) \Big] \mitbf{\hat{e}_1}  \nonumber\\
& + \Big[ -\cos (\theta_p)  \sin (\Omega_c t) \cos (-\Omega_p t) \, + \,  \cos (\Omega_c t)  \sin (-\Omega_p t) \Big] \mitbf{\hat{e}_2}  \nonumber\\
& +\sin (\theta_p)  \sin (\Omega_c t) \mitbf{\hat{e}_3}\, , \\
\mitbf{\hat{e}^p_3} = & \phantom{-}  \, \:  \sin (\theta_p) \Big[  \sin (-\Omega_p t) \mitbf{\hat{e}_1} + \cos (-\Omega_p t) \mitbf{\hat{e}_2}  \Big] + \cos (\theta_p) \mitbf{\hat{e}_3} \, .
\end{align}
\label{eq:eptoe}
\end{subequations}
The reverse relationships, the time-dependent direction of the inertial frame as seen in the mantle frame, is expressed by 

\begin{subequations}
\begin{align}
\mitbf{\hat{e}_1}  = & \phantom{-} \, \:  \Big[  \cos (\theta_p)  \cos (\Omega_c t) \sin (-\Omega_p t) \, - \,  \sin (\Omega_c t)  \cos (-\Omega_p t) \Big] \mitbf{\hat{e}^p_1}  \nonumber\\
&+\Big[- \cos (\theta_p)  \sin (\Omega_c t) \sin (-\Omega_p t) \, - \,  \cos (\Omega_c t)  \cos (-\Omega_p t) \Big] \mitbf{\hat{e}^p_2}  \nonumber\\
& + \sin (\theta_p)  \sin (-\Omega_p t) \mitbf{\hat{e}^p_3} \, ,\\
\mitbf{\hat{e}_2}  =& \phantom{-} \, \: \Big[  \cos (\theta_p)  \cos (\Omega_c t) \cos (-\Omega_p t) \, + \,  \sin (\Omega_c t)  \sin (-\Omega_p t) \Big] \mitbf{\hat{e}^p_1}  \nonumber\\
& + \Big[ -\cos (\theta_p)  \sin (\Omega_c t) \cos (-\Omega_p t) \, + \,  \cos (\Omega_c t)  \sin (-\Omega_p t) \Big] \mitbf{\hat{e}^p_2}  \nonumber\\
& +\sin (\theta_p)  \cos (-\Omega_p t) \mitbf{\hat{e}^p_3} \, ,\\
\mitbf{\hat{e}_3} = & \phantom{-}  \, \:  \sin (\theta_p) \Big[ - \cos (\Omega_c t) \mitbf{\hat{e}^p_1} + \sin (\Omega_c t) \mitbf{\hat{e}^p_2}  \Big] + \cos (\theta_p) \mitbf{\hat{e}^p_3} \, .
\end{align}
\label{eq:etoep}
\end{subequations}

The relationships of Eqs. (\ref{eq:eptoe}-\ref{eq:etoep}) allow one to express any vectorial quantity defined in the inertial frame in its equivalent form as seen in the mantle frame, or vice-versa.  In particular, the rotation vector of the mantle $\mitbf{\Omega}$ is defined in the inertial frame by Eq. (\ref{eq:rot_inertial}).  Using Eqs. (\ref{eq:etoep}), we can express how $\mitbf{\Omega}$ changes as a function of time, as seen in the frame attached to the mantle.  Using standard trigonometric identities, it is straightforward (although somewhat tedious) to show that 

\begin{equation}
\mitbf{\Omega} = \Big[ \Omega_c - \Omega_p \cos (\theta_p) \Big] \mitbf{\hat{e}^p_3} + \Omega_p \sin (\theta_p)  \Big[ \cos (\Omega_c t)  \mitbf{\hat{e}^p_1} - \sin (\Omega_c t) \mitbf{\hat{e}^p_2} \Big] \, . \label{eqa:rot_p}
\end{equation}
Although we have used a different notation, this latter expression is equivalent to Eq. (1) of \citet{eckhardt81} when an exact Cassini state is maintained.
For an observer fixed to the mantle frame, the orientation of the rotation vector $\mitbf{\Omega}$ is offset  from the figure axis $\mitbf{\hat{e}^p_3}$ and precesses about the latter in a retrograde direction at frequency $\Omega_c$.  Let us define $\Omega_o$ as the amplitude of the rotation vector given by

\begin{equation}
\Omega_o = |\mitbf{\Omega} | =  \Big[ \Omega_c^2 + \Omega_p^2 - 2\Omega_c \Omega_p \cos (\theta_p) \Big]^{1/2} \, .
\label{eq:om0a1}
\end{equation}
Since $\Omega_c \gg \Omega_p$, to a good approximation, we can write 

\begin{equation}
\Omega_o \approx \Omega_c - \Omega_p \cos (\theta_p) .
\end{equation}
Defining $\theta_m$ as the angle of offset between $\mitbf{\Omega}$ and $\mitbf{\hat{e}^p_3}$, we can write Eq. (\ref{eqa:rot_p}) as

\begin{equation}
\mitbf{\Omega} = \Omega_o \cos (\theta_m)  \mitbf{\hat{e}^p_3} + \Omega_o \sin (\theta_m)  \mitbf{\hat{e}^p_\perp}(t) \, ,
\end{equation}
where the vector $\mitbf{\hat{e}^p_\perp}(t)$ is given by 

\begin{equation}
\mitbf{\hat{e}^p_\perp} (t) = 
 \Big[ \cos (\omega \Omega_o t)  \mitbf{\hat{e}^p_1} + \sin (\omega \Omega_o t) \mitbf{\hat{e}^p_2} \Big] \, . \label{eqa:eperp}
\end{equation}
and where the frequency $\omega$, expressed in units of cycles per lunar day, is defined as

\begin{equation}
\omega = - \frac{\Omega_c}{\Omega_o} = - 1 - \cos(\theta_p)  \frac{\Omega_p}{\Omega_o} \, .
\end{equation}

The unit vector  $\mitbf{\hat{e}^p_\perp}(t)$ expresses the rotation at frequency $\omega \Omega_o$  of the orientation of $\mitbf{\Omega}$ about $\mitbf{\hat{e}^p_3}$ as seen by an observer in the mantle frame. As $\omega$ is negative, the rotation is retrograde.  Since $\mitbf{\Omega}$ is in the Cassini plane, $\mitbf{\hat{e}^p_\perp}(t)$ describes more generally the retrograde rotation about $\mitbf{\hat{e}^p_3}$ of the longitude of the Cassini plane as seen by an observer in the mantle frame, and is depicted in Figs.~\ref{fig:rotframes}c,d of the main text.  Furthermore, it is easy to show that 

\begin{subequations}
\begin{align}
& \mitbf{\hat{e}^p_3} \times \mitbf{\hat{e}^p_\perp} (t)   = 
 \Big[ - \sin (\omega \Omega_o t) \mitbf{\hat{e}^p_1} + \cos (\omega \Omega_o t)  \mitbf{\hat{e}^p_2} \Big] \, , \\
& \frac{d}{dt} \mitbf{\hat{e}^p_\perp} (t) = \omega \Omega_o \Big[ - \sin (\omega \Omega_o t) \mitbf{\hat{e}^p_1} + \cos (\omega \Omega_o t)  \mitbf{\hat{e}^p_2} \Big] \, , 
\end{align}
where the time derivative is taken in the mantle frame, and therefore we can write

\begin{equation}
\frac{d}{dt} \mitbf{\hat{e}^p_\perp} (t) =  \omega \Omega_o \Big( \mitbf{\hat{e}^p_3} \times \mitbf{\hat{e}^p_\perp} (t) \Big) \, . 
\end{equation}
\label{eq:dteperpapp}
\end{subequations}
Note that the direction of the vector $\mitbf{\hat{e}^p_3} \times \mitbf{\hat{e}^p_\perp} (t)$ is perpendicular to the Cassini plane, towards $\mitbf{\hat{e}^c_2}$ (see Fig.~\ref{fig:rotframes} of the main text).   

The rotation vectors of the fluid core ($\mitbf{\Omega_f}$) and inner core ($\mitbf{\Omega_s}$) can be defined similarly.  They also remain at fixed orientations when viewed in the Cassini frame (see Fig.~\ref{fig:rotframes}b of the main text) and are also precessing at frequency $\omega \Omega_o=-\Omega_c$ when seen by an observer in the mantle frame.      The development used above for the mantle can be used identically for the inner core, with the orientation of the inner core's symmetry axis (with respect to the ecliptic normal) given by $\theta_p + \theta_n$ and the orientation of its rotation vector (with respect to the mantle frame) given by $\theta_m + \theta_s$.  The fluid core does not need to remain in synchronous rotation, but we can represent its rotation rate in a similar manner.  Although it does not have a symmetry axis per say, we can use $\theta_m+\theta_f$ to represent the orientation of both its rotation vector and symmetry axis with respect to the mantle frame to develop an expression for its rotation vector.  The rotation vectors of the fluid core and inner core are then

\begin{subequations}
\begin{align}
\mitbf{\Omega_f} &= \Omega_o^f  \cos (\theta_m +\theta_f)  \mitbf{\hat{e}^p_3} + \Omega_o^f \sin (\theta_m+\theta_f)  \mitbf{\hat{e}^p_\perp}(t) \, ,\\
\mitbf{\Omega_s} &= \Omega_o^s \cos (\theta_m+\theta_s)  \mitbf{\hat{e}^p_3} + \Omega_o^s \sin (\theta_m+\theta_s)  \mitbf{\hat{e}^p_\perp}(t) \, ,
\end{align}
with 

\begin{align}
\Omega_o^f &\approx \Omega_c - \Omega_p \cos (\theta_p+\theta_m+\theta_f) \, ,\\
\Omega_o^s &\approx \Omega_c - \Omega_p \cos (\theta_p + \theta_n) \, .
\end{align}
\end{subequations}
Note that the amplitude of rotation of the mantle, fluid core and inner core are not equal to one another.  However, their amplitude differ by no more than the Poincar\'e number given by the ratio $\Omega_p/\Omega_c = 4.022 \times 10^{-3}$, and except for very large values of $\theta_n$, their difference is typically much smaller than that. Thus, to a good approximation, we can set $\Omega_o^f \approx \Omega_o^s \approx \Omega_o$, in the definition of our rotation vectors, which simplifies the mathematical development of our model.


\section{The Cassini state in the inertial frame} 

As seen in the inertial frame $(\mitbf{\hat{e}_1}, \mitbf{\hat{e}_2}, \mitbf{\hat{e}_3})$ defined in Appendix A, the angular momentum equation describing the rotational dynamics of a single-body Moon is expressed by 

\begin{equation}
\frac{d}{dt} \mitbf{H} = \mitbf{\Gamma}
\end{equation}
where $\mitbf{H}$ is the angular momentum of the whole Moon and $\mitbf{\Gamma}$ is the gravitational torque from Earth.  Assuming a negligible misalignment between the rotation vector and the maximum (polar) principal moment of inertia $C$, we can write $\mitbf{H} = C \mitbf{\Omega}$, where $\mitbf{\Omega}$ is the rotation vector of the single-body Moon, given by Eq. (\ref{eq:rot_inertial}).  Taking the time derivative of $\mitbf{H}$ yields 

\begin{equation}
\frac{d}{dt} \mitbf{H} = - C \Omega_c \Omega_p \sin (\theta_p) \Big[ \cos(-\Omega_p t) \mitbf{\hat{e}_1} - \sin(-\Omega_p t) \mitbf{\hat{e}_2} \Big] \, .\label{eqb:H}
\end{equation}
Focusing, as we do throughout our study, on the long time scale equilibrium, the gravitational torque by Earth averaged over one orbit is in the same direction as the time-derivative of $\mitbf{H}$ and is given by 

\begin{equation}
\mitbf{\Gamma} = - n^2 (\Phi_\beta^p  \bar{A} \beta   + \Phi_\gamma^p  \bar{A} \gamma ) \Big[ \cos(-\Omega_p t) \mitbf{\hat{e}_1} - \sin(-\Omega_p t) \mitbf{\hat{e}_2} \Big] \, ,\label{eqb:tq}
\end{equation}
where we have used Eqs. (\ref{eq:tqearth}) and (\ref{eq:phis}) of the main text, without the inner core contribution.  Setting Eqs. (\ref{eqb:H}) and (\ref{eqb:tq}) equal to one another, we find

\begin{equation}
{C} \frac{\Omega_p \Omega_c}{n^2} \sin (\theta_p) =   \Phi_\beta^p  \bar{A} \beta  +  \Phi_\gamma^p  \bar{A} \gamma  \, .
\end{equation}  
Since $n=\Omega_c-\Omega_p$ and $\Omega_p \ll \Omega_c$, we can approximate $\Omega_c/n \approx 1$, and we 
retrieve (in our notation) the condition on $\theta_p$ given in Eq. (19) of \cite{peale69} that defines the Cassini state of a single body Moon

\begin{equation}
{C} \frac{\Omega_p}{n} \sin (\theta_p) =   \Phi_\beta^p  \bar{A} \beta  +  \Phi_\gamma^p  \bar{A} \gamma  \, .
\end{equation}  

By following a similar procedure, we can construct an expression for the Cassini state of the solid inner core of the Moon.  As seen in the inertial frame, the angular momentum of the inner core ($\mitbf{H_s}$) obeys

\begin{equation}
\frac{d}{dt} \mitbf{H_s} = \mitbf{\Gamma_s}
\end{equation}
where $\mitbf{\Gamma_s}$ is the total torque on the inner core.  Once more assuming a negligible misalignment between the rotation vector $\mitbf{\Omega_s}$ and the maximum (polar) principal moment of inertia $C_s$, we can write $\mitbf{H_s} = C_s \mitbf{\Omega_s}$.  The rotation vector of the inner core is given by an expression analogous to Eq. (\ref{eq:rot_inertial}) but also includes the tilt of the inner core figure $\theta_n$ with respect to the mantle,

\begin{equation} 
\mitbf{\Omega_s} = \Big[ - \Omega_p + \Omega_c \cos (\theta_p+\theta_n) \Big] \mitbf{\hat{e}_3} + \Omega_c \sin (\theta_p+\theta_n)  \Big[ \sin (-\Omega_p t)  \mitbf{\hat{e}_1} + \cos (-\Omega_p t) \mitbf{\hat{e}_2} \Big] \, . \label{eq:rots_inertial}
\end{equation}
Taking the time derivative of $\mitbf{H_s}$ yields 

\begin{equation}
\frac{d}{dt} \mitbf{H_s} = - C_s \Omega_c \Omega_p \sin (\theta_p+\theta_n) \Big[ \cos(-\Omega_p t) \mitbf{\hat{e}_1} - \sin(-\Omega_p t) \mitbf{\hat{e}_2} \Big] \, .\label{eqb:Hs}
\end{equation}
Using Eqs. (\ref{eq:tqs}) and (\ref{eq:phis}), and the approximation $\Omega_o\approx n$, the torque on the inner core is 

\begin{align}
\mitbf{\Gamma_s} = - n^2  \bar{A}_s \Big( & \phantom{-}  \Phi_\beta^n  \beta_s \alpha_3   \, + \, \Phi_\gamma^n \gamma_s \alpha_3 \, +  \, e_s  \alpha_3 \alpha_g  \sin(\theta_n)\cos(\theta_n)  \nonumber\\
& - \,  e_s \alpha_1 \, \sin(\theta_n + \theta_p)\cos(\theta_n + \theta_p )\Big) \,
 \cdot  \Big[ \cos(-\Omega_p t) \mitbf{\hat{e}_1} - \sin(-\Omega_p t) \mitbf{\hat{e}_2} \Big]  \, , \label{eqb:tqs}
\end{align}
where we have assumed $\theta_m+\theta_f=-\theta_p$, the latter corresponding to a fluid core rotation vector aligned with the ecliptic normal.  Setting Eq. (\ref{eqb:Hs}) equal to Eq. (\ref{eqb:tqs}) yields 

 \begin{align}
 \frac{C_s}{\bar{A}_s} & \frac{\Omega_p \Omega_c}{n^2} \sin (\theta_p+\theta_n) = \nonumber\\
&   \Phi_\beta^n  \beta_s \alpha_3  \, + \, \Phi_\gamma^n \gamma_s \alpha_3 \, + \, e_s  \alpha_3 \alpha_g  \sin(\theta_n)\cos(\theta_n) - \, e_s \alpha_1 \, \sin(\theta_n + \theta_p)\cos(\theta_n + \theta_p ) \, .\label{eqb:casssic1}
 \end{align}
The first two terms on the right-hand side of Eq. (\ref{eqb:casssic1}) capture the gravitational torque from Earth averaged over one orbit; they involve products of sines and cosines of $(I+\theta_p+\theta_n)$ (see Eq. \ref{eq:phis}).  The third and fourth terms capture, respectively, the gravitational torque that the rest of the Moon exert on the inner core and the pressure torque from the misaligned rotation vectors of the fluid and solid cores at the ICB. For a given $I$ and $\theta_p$, Eq. (\ref{eqb:casssic1}) gives the condition that the tilt angle $\theta_n$ must obey in order for the inner core to precess about the ecliptic normal at the same rate as the lunar orbit.  In other words, it represents the balance that determines the Cassini state of the inner core of the Moon.  

In the present-day Moon, the internal torque from the mantle and fluid core on the inner core dominates the gravitational torque from Earth. Setting  $\Phi_\beta^n  = \Phi_\gamma^n = 0$ in Eq. (\ref{eqb:casssic1}) and, since $\theta_n$ is typically much larger than $\theta_p=1.543^\circ$ (see Fig.~\ref{fig:ficn}), we can use the following approximation 

 \begin{equation}
 \sin(\theta_n + \theta_p)\cos(\theta_n + \theta_p ) \approx  \sin(\theta_n)\cos(\theta_n) \, ,
 \end{equation}
 which allows to simplify Eq. (\ref{eqb:casssic1}) to 
  
  \begin{equation}
 \frac{C_s}{\bar{A}_s} \frac{\Omega_p}{n} \sin (\theta_p+\theta_n) =  - \omega_{ficn} \sin(\theta_n)\cos(\theta_n)  \, , \label{eqb:casssic2}
 \end{equation}
where $\omega_{ficn}$ is the frequency of the FICN given by Eq. (\ref{eq:omficn}) and where we have removed a factor $\Omega_c/n \approx 1$ on the left-hand side.  Since the dynamical ellipticity of the inner core is small, $C_s \approx \bar{A}_s$, the tilt angle $\theta_n$ in the Cassini state of the inner core depends on the interior structure only insofar as it affects the FICN frequency; different interior density models of the Moon that have the same FICN frequency will have the same tilt angle $\theta_n$.

\acknowledgments
This work was supported by an NSERC/CRSNG Discovery Grant.  Figures were created using the GMT software \cite[]{gmt}.  Comments and suggestions by Rose-Marie Baland and an anonymous reviewer helped to improved this paper.  All data for this paper is properly cited and referred to in the reference list.



\end{document}